\documentclass[superscriptaddress,twocolumn,nobibnotes]{revtex4-2}
\usepackage{amsmath}
\usepackage{amssymb}
\usepackage{bbold}
\usepackage{braket}
\usepackage{graphicx}
\usepackage{color}
\usepackage{stix}
\usepackage[dvipsnames]{xcolor}
\usepackage[colorlinks,linkcolor=RoyalBlue,citecolor=RoyalBlue,urlcolor=magenta]{hyperref}

\begin{document}

\title{CHSH Bell Tests For Optical Hybrid Entanglement}

\author{Morteza~Moradi}
\affiliation{Institute of Informatics, Faculty of Mathematics, Informatics and Mechanics, University of Warsaw, Banacha 2, 02-097 Warsaw, Poland}

\author{Juan Camilo {L\'opez Carre\~no}}
\affiliation{Institute of Theoretical Physics, Faculty of Physics, University of Warsaw, Pasteura 5, 02-093 Warsaw, Poland}
\affiliation{Institute of Informatics, Faculty of Mathematics, Informatics and Mechanics, University of Warsaw, Banacha 2, 02-097 Warsaw, Poland}

\author{Adam Buraczewski}
\affiliation{Institute of Informatics, Faculty of Mathematics, Informatics and Mechanics, University of Warsaw, Banacha 2, 02-097 Warsaw, Poland}

\author{Thomas~McDermott}
\affiliation{Institute of Informatics, Faculty of Mathematics, Informatics and Mechanics, University of Warsaw, Banacha 2, 02-097 Warsaw, Poland}

\author{Beate Elisabeth Asenbeck}
\affiliation{Laboratoire Kastler Brossel, Sorbonne Universit\'e, CNRS, ENS-Universit\'e PSL, Coll\`ege de France, 4 Place Jussieu, 75005 Paris, France}

\author{Julien Laurat}
\affiliation{Laboratoire Kastler Brossel, Sorbonne Universit\'e, CNRS, ENS-Universit\'e PSL, Coll\`ege de France, 4 Place Jussieu, 75005 Paris, France}

\author{Magdalena Stobi\'nska}
\email{magdalena.stobinska@gmail.com}
\thanks{Corresponding author}
\affiliation{Institute of Informatics, Faculty of Mathematics, Informatics and Mechanics, University of Warsaw, Banacha 2, 02-097 Warsaw, Poland}

\date{\today}

\begin{abstract} 
Optical hybrid entanglement can be created between two qubits, one encoded in a single photon and another one in coherent states with opposite phases. It opens the path to a variety of quantum technologies, such as heterogeneous quantum networks, merging continuous and discrete variable encoding, and enabling the transport and interconversion of information. However, reliable characterization of the nature of this entanglement is limited so far to full quantum state tomography. Here, we perform a thorough study of Clauser--Horne--Shimony--Holt (CHSH) Bell inequality tests, enabling practical verification of quantum correlations for optical hybrid entanglement. We show that a practical violation of this inequality is possible with simple photon number on/off measurements if detection efficiencies stay above 82\%. Another approach, based on photon-number parity measurements, requires 94\% efficiency but works well in the limit of higher photon populations. Both tests use no postselection of the measurement outcomes and they are free of the fair-sampling hypothesis. Our proposal paves the way to performing loophole-free tests using feasible experimental tasks such as coherent state interference and photon counting, and to verification of hybrid entanglement in real-world applications.
\end{abstract}

\maketitle

\section{Introduction}

Optical hybrid entanglement is a form of quantum correlations that embodies the original Schr\"odinger's \emph{Gedankenexperiment}~\cite{Schroedinger1935} by replacing the cat with a classical light beam, i.e. a single photon is entangled with a coherent light ~\cite{Jeong2014, Morin2014, Huang2019, Andersen2015}.\break It may become a key asset in resource-efficient quantum computation~\cite{Lee2013}, quantum key distribution~\cite{Rigas2006, Wittmann2010, Yin2019} and quantum buses~\cite{Spiller2006,vanLoock2008}; it has already been employed in complex protocols which paved the way to building heterogeneous quantum networks, e.g., entanglement swapping, and a quantum encoding converter~\cite{Guccione2020, Darras2023}. Furthermore, it was used to probe fundamental limits of quantum theory~\cite{Chen2013, Aspelmeyer2014}, studying hybrid discrete- (DV) and continuous-variable (CV) quantum information~\cite{vanLoock2011, Andersen2015}, and the information capacity of a photonic state~\cite{Nape2017}. 

Quantum technologies often require testing of quantum nonlocality in underlying resources and this is particularly challenging for optical hybrid entanglement. Its dual DV--CV nature implies that Bell nonlocality tests based on hybrid measurement strategies should be optimal. They involve a binary observable measured on the single photon mode and one with a continuous spectrum on the other mode. However, since implementing random Bell test settings in the photon number basis is not experimentally possible, strategies harnessing general qubit measurements are impractical. More universal approaches often employ coarse-graining of detection outcomes for multiphoton entanglement, which can impair Bell nonlocality tests~\cite{Stobinska2011, Stobinska2014}. Moreover, hybrid quantum states decohere exponentially fast with the increasing photon population of the classical light wave~\cite{vanLoock2011, Zurek2003, LeJeannic2018}. Therefore, how to design a nonlocality test for hybrid entanglement that can be set up in a laboratory, remains an open question.

Until now, nonlocality tests based on quantum steering inequality~\cite{Cavailles2018} and several strategies based on the Clauser--Horne--Shimony--Holt (CHSH) Bell inequality~\cite{Clauser1969} have been outlined. In~\cite{Ketterer2016}, the nonlocality of hybrid entanglement was tested with displaced parity measurements on both modes. A hybrid detection strategy was also discussed where the measurement on DV mode was replaced with a general qubit measurement. These tests need detection efficiencies of at least 90\%. In~\cite{Cavailles2019}, two hybrid strategies were considered, that involved either displaced parity or displaced on/off measurements. There, the minimal required detection efficiency is 83\% for the former, and 63\% for the latter test, respectively. However, performing a qubit rotation in the photon-number basis is currently challenging. 

Other optical hybrid Bell tests, e.g. hybrid polarization entanglement~\cite{Kwon2013}, can serve as a guideline. Although this state is physically different from the photon-number entanglement we consider, it shows mathematical similarities. The hybrid nonlocality testing strategies proposed for it involve displaced parity and displaced on/off measurements for the CV mode and a generalized polarization measurement for the DV mode. Albeit they can be implemented with polarizers and photon-number-resolving (PNR) detection of efficiencies higher than 82\%, these Bell tests have not been performed yet.

Here, we perform a thorough analytical and numerical study of two practical CHSH Bell inequality tests possessing a high potential to achieve feasible verification of quantum nonlocality in the optical hybrid entanglement. They can be implemented in an experimental setup, where each mode interferes with a coherent field followed by on/off or parity measurements, realized by means of e.g. PNR detectors~\cite{Gerrits2011}. We show that the first test can achieve the inequality violation for detection efficiencies higher than 82\% and amplitudes of the CV mode below 0.6. In the ideal, lossless conditions, this test would allow for a violation of up to $2.71$. In contrast, the latter measurement scheme works with higher amplitudes but also sets higher requirements for system efficiencies, which must stay above 94\%. This renders it less practical. Finally, for comparison, we also consider hybrid tests, modified for general qubit measurements, and show that in theory, this flavor of hybrid entanglement can maximally violate the CHSH Bell inequality. All the analyzed measurement schemes do not involve postselection and therefore, they have the capability to keep the detection loophole closed.

This paper is organized as follows. Section~\ref{OSCS}  briefly discusses the definition of optical hybrid entanglement as considered in this study as well as its basic properties. In Section~\ref{BNT} we introduce a general Bell test design based on the CHSH Bell inequality, including the experimental setup, model of losses used, and the theoretical background. Next, in Section~\ref{sec:results} we present and discuss numerical results obtained by applying various measurement schemes for the nonlocality tests. Section~\ref{Conc} is devoted to an outlook and conclusions.

\section{Optical hybrid entanglement}\label{OSCS}

Let us consider the following hybrid entanglement that was recently generated~\cite{Morin2014} and subsequently used in various protocols~\cite{Guccione2020, Darras2023, LeJeannic2018a, Parker2020}
\begin{equation}
\ket{\Psi} = \frac{1}{\sqrt{2}} \bigl( \ket{0}_A\ket{\text{cat}^-}_B + \ket{1}_A \ket{\text{cat}^+}_B \bigr). \label{HEC}
\end{equation}
In mode $A$, a discrete-variable qubit is encoded in photon-number states $\ket{0}$ and $\ket{1}$ -- the vacuum and single-photon Fock state, respectively. The qubit is entangled with mode $B$ that carries two mutually orthogonal continuous-variable states, $\ket{\text{cat}^-}$ and $\ket{\text{cat}^+}$. Both of them are superpositions of coherent states of the same amplitude but opposite phases
\begin{equation}
\ket{\text{cat}^\pm} = \frac{1}{N_\pm} \bigl( \ket{\gamma} \pm \ket{-\gamma} \bigr),
\label{cat}
\end{equation}
where $N_\pm = \sqrt{2(1\pm e^{-2|\gamma|^2})}$ is the normalization constant and $\ket{\gamma} = e^{-\frac{\lvert\gamma\rvert^2}{2}}\sum_{n=0}^{\infty} \frac{\gamma^n}{\sqrt{n!}}|n\rangle$ is a coherent state. States $\ket{\text{cat}^\pm}$ are orthogonal, $\braket{\text{cat}^+\vert \text{cat}^-}=0$, which stems from the fact that $\ket{\text{cat}^+}$ carries only even photon-number components while $\ket{\text{cat}^-}$ only odd ones. This property works as a hint that PNR-based measurement schemes could be the best suited for their discrimination.

Full characterization of $\ket{\Psi}$ was realized using efficient quantum state tomography~\cite{Morin2014} and quantum steering tests were also performed ~\cite{Cavailles2018}. However, a full nonlocality test has not been carried out yet. Interestingly, in the limit of $\gamma \to 0$, $\ket{\Psi}$ tends to the single-photon entanglement $\frac{1}{\sqrt{2}} \bigl(\ket{0}\ket{1}-\ket{1}\ket{0}\bigr)$. Also in this case verification of nonlocality and entanglement in $\ket{\Psi}$ is conceptually challenging and it has generated a lot of attention~\cite{Tan1991, Cooper2008, Das2021, Das2022a, Das2022b, Das2022c}.

\section{General Bell test design for optical hybrid entanglement}
\label{BNT}

\begin{figure}[t]\centering
\includegraphics[width=\columnwidth]{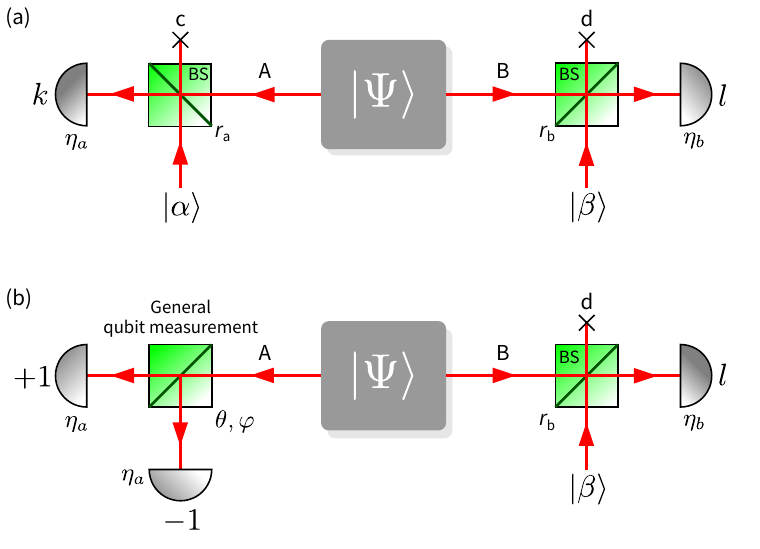}
    \caption{Nonlocality Bell tests for an optical hybrid bipartite photon-number entanglement $\ket{\Psi}$, as defined in Eq.~(\ref{HEC}). (a) Two modes of the state are locally interfered with coherent fields, $\ket{\alpha}$ and $\ket{\beta}$, followed by measurements of efficiency $\eta_a$ and $\eta_b$. Here, the variable beam splitters' reflectivities $r_a$ and $r_b$ act as the Bell test settings. The registered readouts $k$ and $l$ are coarse-grained into the binary outcomes of on/off or even/odd measurements used in the CHSH Bell inequality (\ref{CHSH}).\break (b) Tests with hybrid measurement strategy, where a measurement on mode $A$ is replaced by a general qubit measurement.}
    \label{scheme}
\end{figure}

Verification of quantum nonlocality requires designing a viable Bell test, a task that is quantum state-specific~\cite{Horodecki2009, Brunner2014}. However, universal methods for finding a well-matched one are not known. Moreover, experimental implementations of Bell tests are limited by the number of accessible measurement strategies, which in the case of photonic setups comprise polarization rotations, displacement operations, homodyne, and detectors of high efficiency, e.g., superconducting nanowires~\cite{Reddy2020} or transition-edge sensors~\cite{Lita2008}. Furthermore, within the current state of the art, the implementation of rotations in the Fock state basis, which will result in the creation of superpositions of states with different photon numbers, is challenging.

The CHSH Bell inequality, which facilitated the first loophole-free Bell tests ~\cite{Giustina2013, Giustina2015}, takes the form
\begin{equation}
    S = \langle A_1 B_1 \rangle + \langle A_1 B_2 \rangle + \langle A_2 B_1 \rangle - \langle A_2 B_2 \rangle \leq 2,
    \label{CHSH}
\end{equation}
where $A_i$ ($B_i$), for $i=1,2$, are binary observables of values $\pm 1$, measured on mode $A$ ($B$). It was successful because of its robustness against noise and errors in experimental settings. Quantum correlations allow $S$ to achieve values up to $2\sqrt{2} \approx 2.83$ and subsequently violate inequality~(\ref{CHSH}). Thus, observing CHSH values clearly crossing the classical boundary of 2 is the goal of any practical test.

We propose to employ the CHSH inequality and either on/off or parity measurements for testing optical hybrid entanglement. In Fig.~\ref{scheme}(a), two modes of the shared state $\ket{\Psi}$ are locally interfered with coherent fields, $\ket{\alpha}$ and $\ket{\beta}$, followed by measurements of efficiency $0<\eta_{a,b}\leq 1$. The variable beam splitters' reflectivities $r_a$ and $r_b$ act as the Bell test settings. The registered readouts $k$ and $l$ are then coarse-grained into two sets, either\ zero/non-zero or even/odd numbers of photons (Sections \ref{ZeroT} and \ref{EOT}).

This design builds on a group theory result that maps two-mode Fock states $|n\rangle |\Sigma-n\rangle$ to Dicke spin-$\frac{\Sigma}{2}$ states with component $S_z\!=\!\frac{\Sigma}{2}\!-\!n$, by means of the Schwinger representation~\cite{Chruscinski}. Next, we observe that any arbitrary rotation of\break $S_z$-spin component encoded in a product of Fock states is easily implemented by means of Fock state interference on a beam splitter~\cite{Kravchuk}. Reflectivities set the spin rotation angles. We also note the fact that testing photon-number correlations requires erasing `how-many-photons' information before taking measurements. This step requires that we locally interfere each mode of the photon-number entangled state with a quantum superposition of indefinite number of photons, e.g., a coherent state, not merely with a Fock state. Effectively, if we examine this interference in the Fock state basis, it will amount not only to the spin rotation, but also to varying the spin value $\frac{\Sigma}{2}$ each time the measurements are taken. In this way, in the Bell test, we no longer need to perform rotations directly in the Fock state basis, but we carry out rotations of a spin that fluctuates in length instead. These Bell tests can detect nonlocality in a wide class of bipartite entanglement: squeezed vacuum, single-photon entanglement, entangled coherent states, and generalized Holland-Burnett states~\cite{Banaszek1999, Wodkiewicz2000, Kuzmich2001, Jeong2003, Lee2011, Ketterer2016, Mycroft2023}. 

In Fig.~\ref{scheme}(b) we further modify the setup from Fig.~\ref{scheme}(a)  for generalized qubit measurements in mode $A$. The goal is to see how much improvement such a test can offer (Sections \ref{HZeroT} and \ref{HEOT}). To this end, we assume that the above-mentioned rotations are experimentally feasible. 

To investigate the collective effect of losses in the state generation, transmission, and imperfect detection in full optical tests, we model them with a beam splitter of transmitivity $\eta_{a,b}$ inserted in front of each detector. In hybrid Bell tests, we avoid the post-selection loophole by assigning the value +1 to the measurement outcome of observable $A_i$ if a `no-detection' event in mode $A$ occurs, i.e. the effective observable could be written as $A^{\text{eff}}_i=\eta A_i + (1-\eta) \mathbb{1} $. 

We also notice that for the homodyne limit (HD), characterized by very small reflectivities, $r_a, r_b \to 0$, and large amplitudes of the coherent fields, $|\alpha|^2, |\beta|^2 \to \infty$, the beam splitter interference may be approximated with displacement operators $D(\delta_{\alpha,\beta})$ with $\delta_\alpha = -i \alpha \sqrt{r_a}$ and $\delta_\beta = -i \beta \sqrt{r_b}$~\cite{Sekatski2012, Paris1996}. In this case, these displacements become the measurement settings. However, in this limit, the photon population of the local oscillators is several orders of magnitude larger than that of the measured state, $\lvert\beta\rvert^2\gg \lvert\gamma\rvert^2$, which may pose additional experimental challenges, e.g.the PNR detection of large photon numbers~\cite{Sekatski2012, Lvovsky2009}.

Details of analytical derivations and numerical methods used for obtaining the results presented in Sections \ref{ZeroT}--\ref{HEOT} are given in the Supplementary Material.

\section{Results}
\label{sec:results}

\subsection{CHSH test with on/off measurements}\label{ZeroT}

The first CHSH Bell test that we will outline is based on the scheme shown in Fig.~\ref{scheme}(a), where the detection outcomes $k$ and $l$ are split into four sets according to the vacuum and non-vacuum events registered in modes $A$ and $B$,\break $\{(k\!\!=\!\!0, l\!\!=\!\!0), (k\!\!=\!\!0, l\!\!\not=\!\!0), (k\!\!\not=\!\!0, l\!\!=\!\!0), (k\!\!\not=\!\!0, l\!\!\not=\!\!0)\}$;\break they correspond to the binary measurement outcomes $\{(+1,+1), (+1,-1), (-1,+1), (-1,-1)\}$ assigned to $A$ and $B$, respectively. The on/off measurement operator takes the form
\begin{equation}
  \label{eq:onoffmeasurement}
  \Pi_{\text{on/off}} = \ket{0}\bra{0} - \sum_{n=1}^{\infty} \ket{n}\bra{n} = 2\ket{0}\bra{0}-\mathbb{1},
\end{equation}
where $\mathbb{1}$ denotes the identity operator. We would like to emphasize that $\Pi_{\text{on/off}}$ describes solely the detection, not the observable employed in the Bell test. It is challenging to concisely analytically express a single observable $A_i$ (or $B_j$) and to reveal its dependence on the settings of the Bell test ($r_a$ and $r_b$), but we can write down explicitly the correlation function between these observables as follows
\begin{equation}
     \langle A_i\, B_j\rangle = \begin{aligned}[t]
     \mathrm{Tr} \Bigl \lbrace & \Pi_{\text{on/off}}^{(A)}
      \Pi_{\text{on/off}}^{(B)}
      \mathrm{Tr}_{c,d}\bigl[ \mathcal{U}_{\text{BS}}(r_{a_i})\,
      \mathcal{U}_{\text{BS}}(r_{b_j}) 
      \\
      &\rho_\Psi\,\rho_\alpha\,\rho_\beta
      \\
      &\mathcal{U}_{\text{BS}}^\dagger (r_{b_j})
      \mathcal{U}_{\text{BS}}^\dagger (r_{a_i}) \bigr]
    \Pi_{\text{on/off}}^{(B)^\dagger}
      \Pi_{\text{on/off}}^{(A)^\dagger}\Bigr \rbrace, 
      \end{aligned}
      \label{eq:onoffcorr}
\end{equation}
where $\rho_\Psi=\ket{\Psi}\bra{\Psi}$, $\rho_\alpha$ and $\rho_\beta$ are the coherent states $\ket{\alpha}$ and $\ket{\beta}$. The action of a beam splitter is described by the unitary $\mathcal{U}_{\text{BS}}(r)=e^{-i\mu(a^\dagger b - a b^\dagger)}$, where $\mu=2\arcsin\sqrt{r}$, $r$ is the beam splitter reflectivity and $a,b$ denote the annihilation operators of interfering modes.

\begin{figure}[t]
    \centering
    \includegraphics[width=\linewidth]{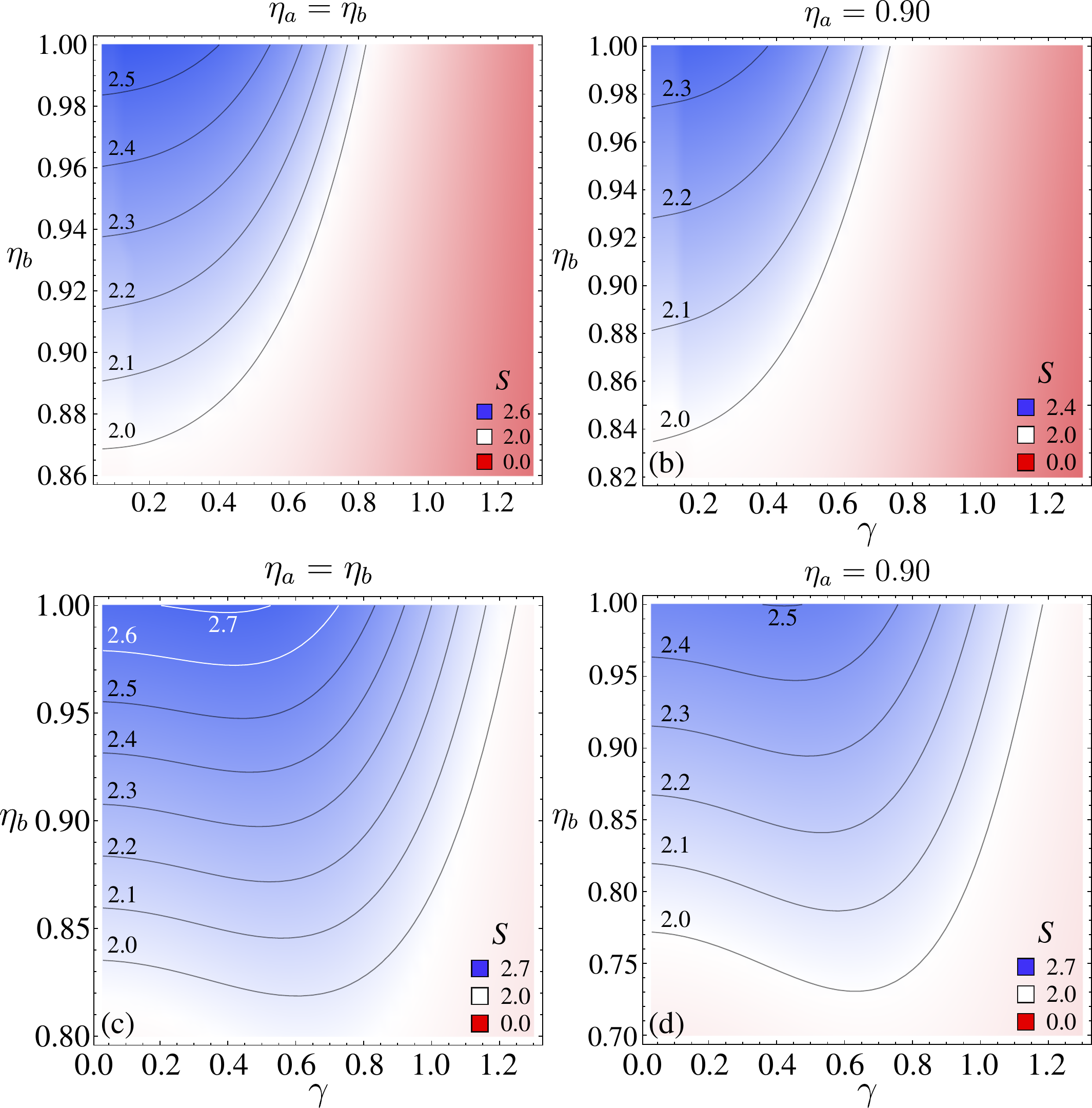}
    \caption{Violation of the CHSH inequality with on/off measurements computed for the hybrid entanglement $\lvert \Psi\rangle$, as a function of the amplitude~$\gamma$ of the state~$\ket{\mathrm{cat}^\pm}$ and (a)~equal detection efficiencies in both arms, $\eta_\mathrm{a} = \eta_\mathrm{b}$ and $\alpha=\beta=2$, and (b)~fixed qubit detection efficiency, $\eta_\mathrm{a} = 0.9$ and $\alpha=\beta=2$.  Panels (c) and (d) display the counterparts of panels (a) and (b), respectively, but in the homodyne limit.}
  \label{QBell-on-off}
\end{figure}

\begin{figure}[t]
    \centering
    \includegraphics[width=\linewidth]{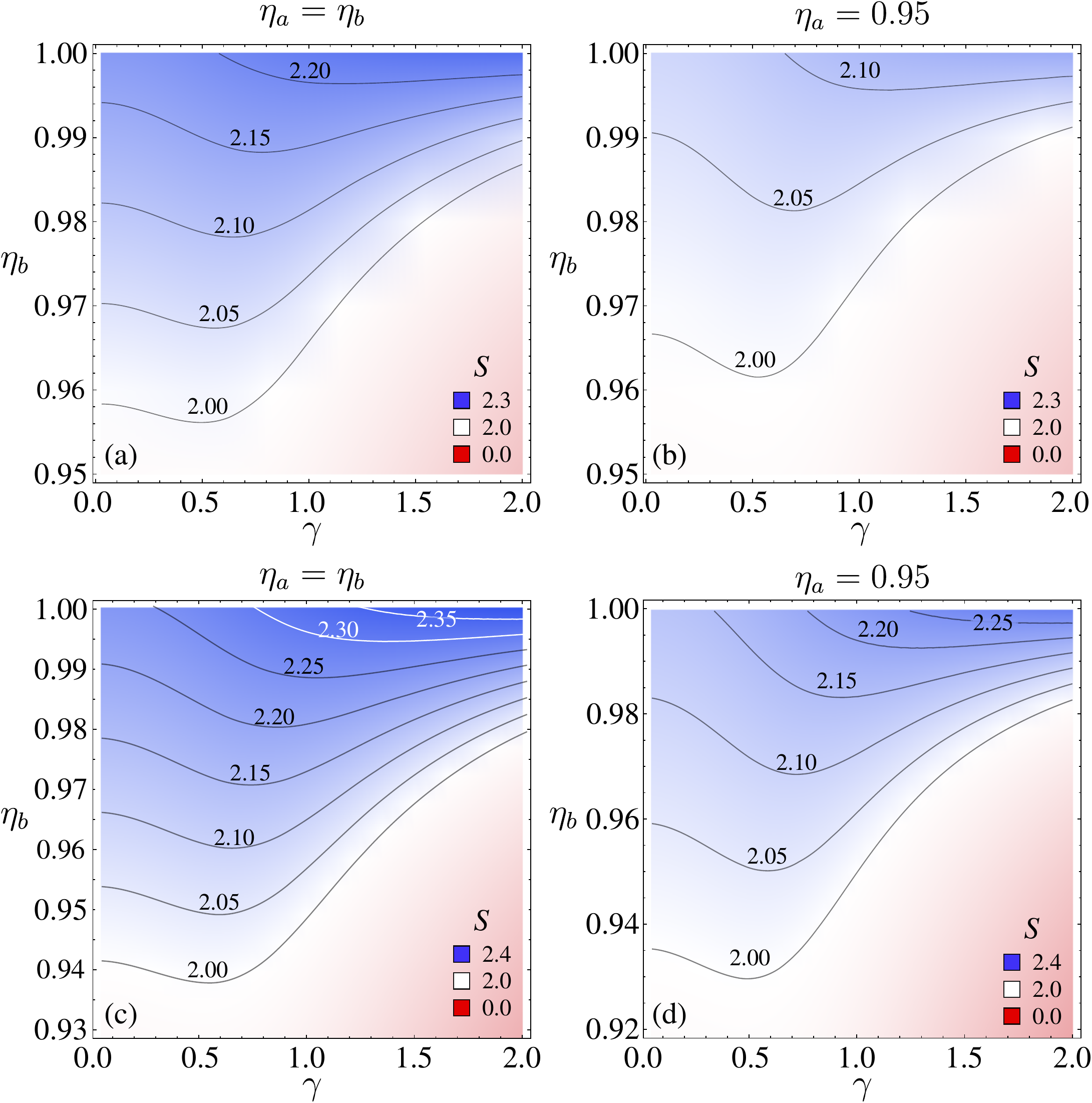}
    \caption{Violation of the CHSH inequality with parity measurements computed for the hybrid entanglement $\lvert \Psi\rangle$, as a function of the amplitude~$\gamma$ of the state~$\ket{\mathrm{cat}^\pm}$ and (a)~equal detection efficiencies in both arms, $\eta_\mathrm{a} = \eta_\mathrm{b}$ and $\alpha=\beta=1$, and (b)~fixed qubit detection efficiency and $\alpha=\beta=1$, $\eta_\mathrm{a} = 0.95$. Panels (c) and (d) display the counterparts of panels (a) and (b), respectively, but in the homodyne limit.}
  \label{QBell-parity}
\end{figure}

To compute the maximal value of parameter $S$ and achieve the CHSH inequality violation for given $\gamma$ and detection efficiencies $\eta_{a,b}$, we numerically optimize Eq.~(\ref{CHSH}) with the correlation functions given by Eq.~(\ref{eq:onoffcorr}), with respect to the Bell test settings $r_{a_{1,2}},r_{b_{1,2}}$ for given parameters $\alpha,\beta$. Next, we optimize the amplitudes $\alpha,\beta$ to obtain the minimal detection efficiencies $\eta_{a,b}$ required for the violation. Numerical optimization is simplified by using the real domain for parameters $\gamma$, $\alpha$ and $\beta$. This approach is viable as phases can be aligned in the experimental implementation.

The results for the on/off measurements are depicted in Fig.~\ref{QBell-on-off}. In panel~(a) we fix the parameters~$\alpha\!=\!\beta\!=\!2$ and set equal detection efficiencies $\eta_a\!=\!\eta_b$. In the limit $\gamma \to 0$ and perfect detection ($\eta_{a,b}=1$) we obtain $S=2.57$. The value of $S$ gradually drops for higher $\gamma$ and reaches the boundary $S=2$ for $\gamma=0.81$, because the odds of spotting the zero-photon measurement quickly vanishes for highly-occupied states. Observing Bell inequality violation becomes also impossible below $\eta_a\!=\!\eta_b=0.87$. For comparison, when we set higher~$\alpha\!=\!\beta\!=\!7$, in the limit $\gamma \to 0$ and perfect detection we obtain maximum violation of $S=2.69$, the same value we got in Ref.~\cite{Mycroft2023} for the single-photon case.

Next, we studied the interplay between the minimal requirements for unequal detection efficiencies. Fig.~\ref{QBell-on-off}(b) shows the violations achieved for $\eta_a\!=\!0.9$ and $\alpha\!=\!\beta\!=\!2$. They are observed for $\gamma\!<\!0.75$ and $\eta_b\!>\!0.83$. This gain comes at the expense of lowering the value of maximal violation to $S\!=\!2.35$. 

The computation results for the homodyne limit are shown in the lower row of Fig.~\ref{QBell-on-off}. We first derived an analytical formula for correlation functions and then numerically maximized the $S$ with respect to the Bell test settings~$\delta_{\alpha_{1,2}},\delta_{\beta_{1,2}}$. We found that in this case, in panel~(c), the CHSH violation is higher than in panel~(a), and it is also observed for a wider span of~$\gamma$. Furthermore, the maximum value of the violation increases to $S=2.71$ for $\gamma=0.4$ and for the settings $\delta_{\alpha_1}=0.18 , \hspace{1mm}\delta_{\alpha_2}=-0.56 , \hspace{1mm} \delta_{\beta_1}=0.17 , \hspace{1mm}\delta_{\beta_2}=-0.61 $, and the minimum efficiency that still provides violation is $\eta_{a,b}=0.82$ for $\gamma=0.6$. Fig.~\ref{QBell-on-off}(d) displays the CHSH violation for $\eta_a=0.9$ and it is the counterpart of panel~(b) in the homodyne limit. Here, the maximal achieved value is higher than in panel~(b), reaching $S=2.5$ for $\gamma=0.42$, and the minimal efficiency that still provides violation gets to $\eta_b=0.73$ for $\gamma=0.63$.

In summary, our results show that significant CHSH violations can be achieved for hybrid entanglement when performing on/off measurements. Notably, such violations can be reached even when the intensity of the local coherent state is comparable to the photonic occupation of the state under study. We also calculated how $S$ increases in the homodyne limit and showed that we can reach a higher violation in this limit.

\subsection{CHSH test with parity measurements}\label{EOT}

Let us now perform coarse-graining of the measurement results from Fig.~\ref{scheme}(a) by assigning $-1$ to the detection of even numbers of photons ($k = \textrm{even}$ or $l=\textrm{even}$), and $+1$ to the detection of odd numbers of photons ($k = \textrm{odd}$ or $l=\textrm{odd}$). This strategy mimics the use of parity measurements on modes $A$ and $B$ with
\begin{equation}
    \Pi_{\text{parity}} = \sum_{n=0}^{\infty} (-1)^{n} \ket{n}\bra{n}\,.
    \label{eq:paritymeasurement}
\end{equation}
Following the steps described in Section~\ref{ZeroT}, we numerically optimize the CHSH value $S$ in Eq.~(\ref{CHSH}) with correlations computed using Eq.~(\ref{eq:onoffcorr}) but replacing the operator $\Pi_{\text{on/off}}$ with $\Pi_{\text{parity}}$ from Eq.~(\ref{eq:paritymeasurement}).

The results for the parity measurement are shown in Fig.~\ref{QBell-parity}. Panel~(a) depicts computations for equal efficiences $\eta_a\!=\!\eta_b$. Here $\alpha\!=\!\beta\!=\!1$ provide us with the smallest $\eta_{a,b}\!=\!0.96$ for which we see the violations. We find that the maximal CHSH value increases from $S\!=\!2.17$ for $\gamma\to0$ to $S\!=\!2.24$ for $\gamma\!=\!2$. We note that these measurements allow to observe violation for larger amplitudes of~$\gamma$ than with on/off measurements. In fact, with this measurement, one could get $S\!\geq\!2$ for $\eta \to 1$ independently of $\gamma$, unlike for the on/off case, where $\gamma$ was limited. However, parity measurements in general require higher detection efficiencies, because losing a photon can turn an even number of photons into an odd one, and vice versa.

Fig.~\ref{QBell-parity}(b) shows computations for $\eta_a\!=\!0.95$ and $\alpha\!=\!\beta\!=\!1$. This efficiency is almost at the minimum necessary for the CHSH violation. Here, $S\!=\!2.14$ is reached for $\gamma\!=\!2$. A value of $\eta_b$ required to witness any violation is $\eta_b\!=\!0.96$ for $\gamma\!=\!0.5$. 

\begin{figure}[t]
    \centering
    \includegraphics[width=\linewidth]{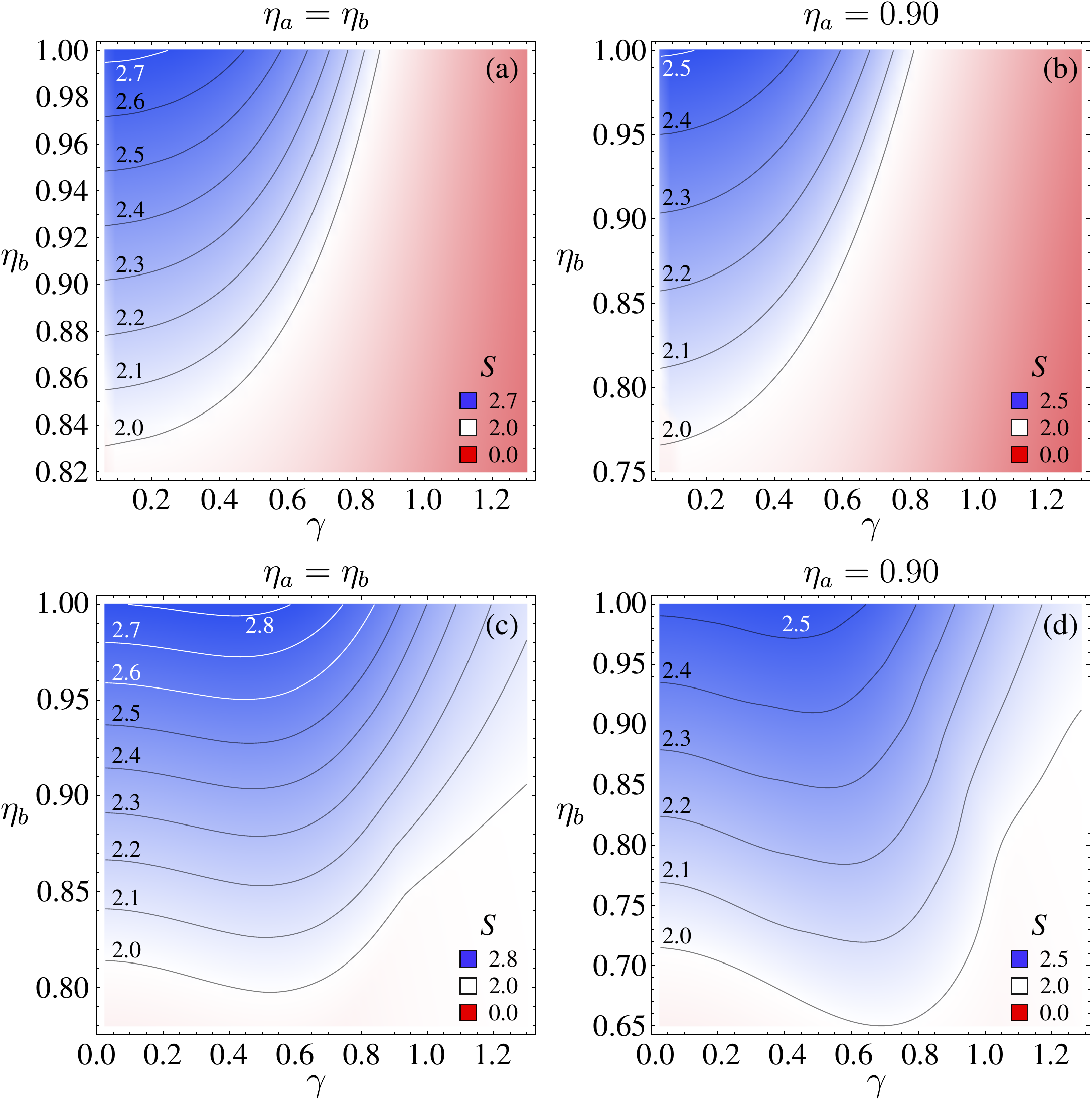}
    \caption{Violation of the hybrid CHSH Bell test with on/off measurements computed for the hybrid entanglement $\lvert \Psi\rangle$, as a function of the amplitude~$\gamma$ of the state~$\ket{\mathrm{cat}^\pm}$ and (a)~equal detection efficiencies in both arms, $\eta_\mathrm{a} = \eta_\mathrm{b}$ and $\alpha=\beta=2$, and (b)~fixed qubit detection efficiency, $\eta_\mathrm{a} = 0.9$ and $\alpha=\beta=2$. Panels (c) and (d) display the counterparts of panels (a) and (b), respectively, but in the homodyne limit.}
  \label{fig:hybridonoff}
\end{figure}
\begin{figure}[t]
    \centering
    \includegraphics[width=\linewidth]{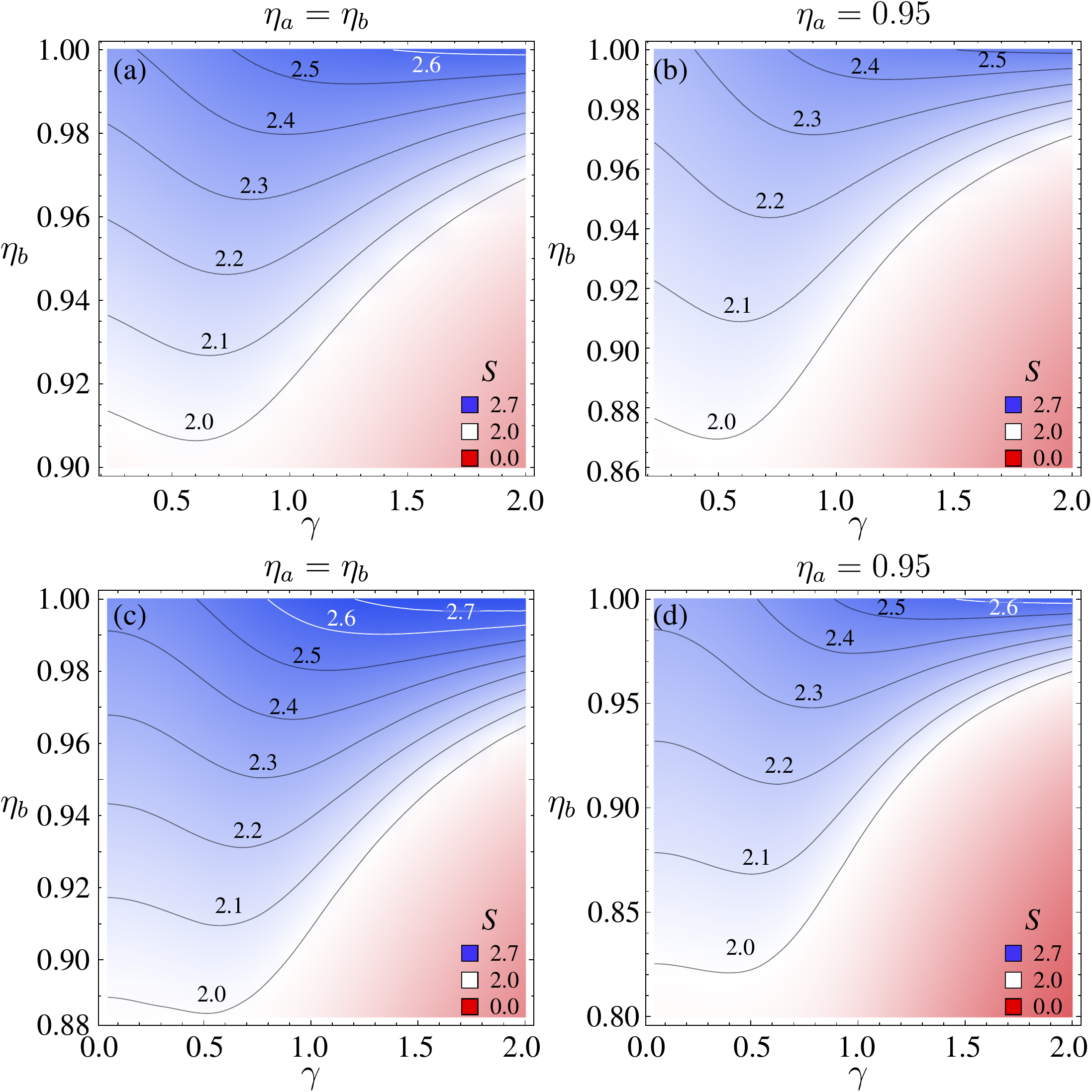}
    \caption{Violation of the hybrid CHSH Bell test with parity measurements computed for the hybrid entanglement $\lvert \Psi\rangle$, as a function of the amplitude~$\gamma$ of the state~$\ket{\mathrm{cat}^\pm}$ and (a)~equal detection efficiencies in both arms, $\eta_\mathrm{a} = \eta_\mathrm{b}$ and $\alpha=\beta=1$, and (b)~fixed qubit detection efficiency, $\eta_\mathrm{a} = 0.95$ and $\alpha=\beta=1$. Panels (c) and (d) display the counterparts of panels (a) and (b), respectively, but in the homodyne limit.}
  \label{fig:hybridparity}
\end{figure}

In the homodyne limit, the results are shown in panels (c) and (d). For $\eta_a\!=\!\eta_b$, panel (c), the violation of $S\!=\!2.39$ is achieved for $\gamma\!=\!2$ and it will increase to maximum violation of $S\!=\!2.44$ for $\gamma\to \infty$. However, it very quickly degrades with losses. The lowest efficiency for which $S > 2$ is $\eta_{a,b}\!=\!0.94$ for $\gamma\!=\!0.54$. In the case of $\eta_a=0.95$, panel (d), the violation of $S\!=\!2.29$ attained for $\gamma\!=\!2$, and $S > 2$ is reached for minimum $\eta_b = 0.93$ and $\gamma\!=\!0.5$.

Thus, our results show that CHSH violations can also be achieved for hybrid entanglement when performing parity measurements. The magnitudes of the violations are lower than in the on/off case, and require better detection efficiencies, but can be observed at larger values of the amplitude~$\gamma$. In both cases, the ideal setting which gave us the maximum violation happens in the homodyne limit.

\subsection{Hybrid CHSH test with on/off measurements}\label{HZeroT}

Another strategy linked to the specific nature of the considered state is to perform the hybrid Bell test shown in Fig.~\ref{scheme}(b). In this design, a general qubit measurement is performed on the single-photon subsystem, while on/off measurements, Eq.~(\ref{eq:onoffmeasurement}), are applied to mode $B$. Every qubit observable can be described in terms of Pauli operators $\sigma_X$, $\sigma_Y$, and $\sigma_Z$~\cite{Nielsen2010}
\begin{equation}
	A_j = \sin \theta_j \cos \phi_j \, \sigma_X + \sin \theta_j \sin \phi_j\, \sigma_Y + \cos \theta_j\, \sigma_Z,
 \label{canonical}
\end{equation}
where $j=1,2$ correspond to two measurement settings used in mode $A$ in the test.

We follow the steps described in Section~\ref{ZeroT} for the optimization of the CHSH value with the on/off measurements, but we apply the observable defined Eq.~(\ref{canonical}) in mode $A$ to achieve hybrid inequality. The results are depicted in Fig.~\ref{fig:hybridonoff}. Panel (a) is computed for $\alpha\!=\!\beta\!=\!2$ and $\eta_a\!=\!\eta_b$. Here, $S\!=\!2.72$ is achieved for $\gamma\to0$, but CHSH violation is reached only for $\eta_{a,b}\!\geq\!0.82$. The range for which the violation can be observed is bounded by $\gamma\!=\!0.88$. Panel (b) displays $\eta_a=0.9$ for $\alpha\!=\!\beta\!=\!2$, for which the maximal violation is $S\!=\!2.54$ but the minimal efficiency needed for the violation drops to $\eta_b\!=\!0.77$. 

In the homodyne limit, the minimal required detection efficiency to observe any CHSH inequality violation is $\eta_{a,b}\!=\!0.8$, Fig.~\ref{fig:hybridonoff}(c). The highest violation of $S\!=\!2.83$ is obtained for $\gamma={\sqrt{\ln 2}}/{2} \approx 0.42$. For $\eta_a\!=\!0.9$, panel (d), the minimal $\eta_b$ equals $0.65$ and maximal violation is $S=2.55$. 

The numerical result from panel (c) can be additionally shown analytically using the CHSH rigidity theorem, which states that if a quantum state maximally violates the CHSH inequality, it is equivalent to the maximally entangled Bell pair $\frac{1}{\sqrt{2}}\bigl(\ket{0}\ket{0}+\ket{1}\ket{1}\bigr)$, while the measurements used are equivalent to the canonical qubit measurements $A_1 = \sigma_Z,\hspace{1mm} A_2 = \sigma_X,\hspace{1mm} B_1 = \frac{1}{\sqrt{2}}(\sigma_Z+\sigma_X), \hspace{1mm} B_2 = \frac{1}{\sqrt{2}}(\sigma_Z-\sigma_X)$~\cite{Summers1987,Gheorghiu2017}. This prerequisite is fulfilled by $\ket{\Psi}$, because Eq.~(\ref{HEC}) already possesses the form of a Bell pair. Orthonormal states $\ket{e_0} = \ket{\text{cat}^-}$ and  $\ket{e_1} = \ket{\text{cat}^+}$ can be considered as two states of a multiphoton qubit, but they must be complemented by an arbitrary set of orthonormal $\ket{e_i}$ where $i \geq 2$ to form the full basis in an infinite-dimensional Hilbert space. 

In order to analytically investigate this theorem, let us take $A_1 = \sigma_Z$ and $A_2 = \sigma_X$, as well as $\delta_{\beta_{1,2}}=\pm\gamma$. With regards to the measurement operators acting on mode $B$, we note that in the homodyne limit they can be expressed as $B_j = D(\delta_{\beta_j})\Pi_{\text{on/off}} D^\dagger(\delta_{\beta_j}) = 2\ket{\delta_{\beta_j}}\bra{\delta_{\beta_j}}-\mathbb{1}$, $j=1,2$. In the basis $\ket{e_i}$, they have the following form
\begin{subequations}
    \begin{align}
    B_j =& \tilde{B}_j \oplus -\mathbb{1},\\
    \tilde{B}_j =& (-1)^{j-1} \sqrt{1-e^{-4\gamma^2}}\, \sigma_X + e^{-2\gamma^2} \, \sigma_Z,
    \end{align}
\end{subequations}
 where $j=1,2$. The Supplementary Material details the full proof.
 
Now, when we substitute $\gamma = \frac{1}{2}\sqrt{\ln 2}\approx 0.42$, then $B_1\!=\!\frac{1}{\sqrt{2}}(\sigma_Z+\sigma_X)$ and $B_2\!=\!\frac{1}{\sqrt{2}}(\sigma_Z-\sigma_X)$ and the maximal violation of the CHSH inequality $S=2\sqrt{2}$ can be observed, thus proving the CHSH rigidity for the optical hybrid entanglement. This stays in agreement with Ref.~\cite{Dastidar2022}.

In summary, the hybrid Bell test, in which a general qubit measurement is performed on mode $A$ while mode $B$ is tested with on/off measurements, allows one to achieve higher CHSH inequality violations compared to the case where on/off measurements are performed on both modes. Furthermore, in the homodyne limit, this approach allows one to observe the maximal CHSH violation of $2\sqrt{2}$ for the hybrid entanglement, showing that this state is equivalent to the maximally entangled Bell pair.

\subsection{Hybrid CHSH test with parity measurements}\label{HEOT}

Lastly, we consider a hybrid test in which general qubit measurements, Eq.~(\ref{canonical}), are performed on mode $A$, and the parity measurements, Eq.~(\ref{eq:paritymeasurement}), on mode $B$. Numerical computations follow the steps from Section~\ref{HZeroT}.

The results are displayed in Fig.~\ref{fig:hybridparity}. Panel (a) is computed for $\alpha\!=\!\beta\!=\!1$ and $\eta_a\!=\!\eta_b$. Here, the violation reaches $S\!=\!2.62$ for $\gamma\!=\!2$ and drops to $S\!=\!2$ for $\eta_{a,b}=0.91$ and $\gamma\!=\!0.6$. In the case of $\eta_b=0.95$, computed for $\alpha\!=\!\beta\!=\!1$, panel (b), maximal violation of $S\!=\!2.52$ is reached for $\gamma\!=\!2$. The CHSH inequality is minimally violated for $\eta_b\!=\!0.87$ for $\gamma\!=\!0.45$.

In the homodyne limit, the CHSH parameter computed for $\eta_a\!=\!\eta_b$ reaches $S\!=\!2.77$ for $\gamma\!=\!2$, panel (c). The minimum required efficiency is $\eta_a\!=\!\eta_b\!>\!0.88$ for $\gamma\!=\!0.47$. For $\eta_a\!=\!0.95$, panel (d), minimal $\eta_b \!>\! 0.83$ for $\gamma\!=\!0.37$, while $S$ takes the maximal value of $2.69$ for $\gamma \to \infty$.
The CHSH inequality in this case is maximally violated for $\gamma \to \infty$ and $\eta_{a,b}\to 1$, which can be also proved analytically. In this limit, $\braket{-\gamma|\gamma} = 0$ and observables $B_1$ and $B_2$ become
\begin{equation}
     B_j = e^{-2|\delta_j|^2} ( 
     \cos\varphi_j \sigma_Z  - \sin \varphi_j \sigma_Y  )\,,
\end{equation}
where~$\varphi_j =4\gamma\,\textrm{Im}(\delta_j)$ and $j=1,2$. Then, by substituting $\delta_{\beta_{1,2}} = \pm i \pi / 16\gamma$ one gets $B_1 = \frac{1}{\sqrt{2}}(\sigma_Z-\sigma_Y)$ and $B_2 = \frac{1}{\sqrt{2}}(\sigma_Z+\sigma_Y)$, which are the canonical measurements up to a rotation in the X--Y plane.
The CHSH inequality can be maximally violated, $S = 2\sqrt{2}$, for $A_1 = \sigma_Z, A_2 = \sigma_Y$.

To sum up this case, in analogy to Section \ref{HZeroT}, the hybrid CHSH test with parity measurements allows one to achieve higher CHSH Bell inequality violations than the test using parity measurements in both modes, at the expense of the feasibility of the experimental scheme. It also allows one to obtain the maximal violation for a specific set of system parameters.

\section{Discussion and conclusion}
\label{Conc}

\begin{figure}
    \centering
    \includegraphics[width=\linewidth]{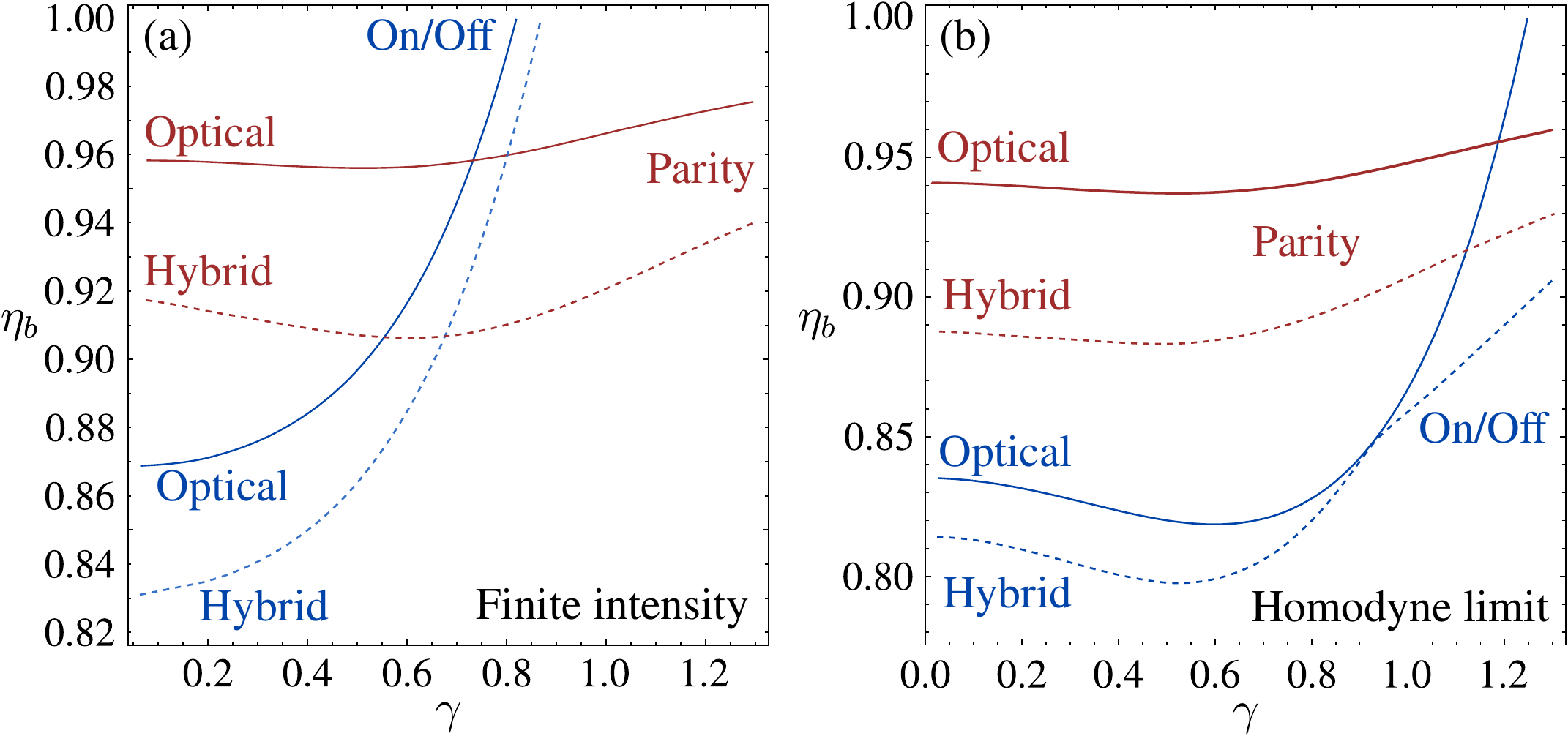}
    \caption{Comparison of nonlocality CHSH Bell tests for optical photon-number hybrid entanglement, Eq.~(\ref{HEC}), with various detection strategies: on/off (blue lines) and parity (red line) with full optical (solid lines) and hybrid (dashed lines) measurements. We assume all the measurement devices to be equally efficient, $\eta_a=\eta_b$. The various lines correspond to the thresholds above which CHSH violation is observed when the local interferences (a)~have finite intensity and (b)~are in the homodyne limit; i.e., the lines are the~$S=2$ isolines from panels~(a) and panels~(c), respectively, of Figs.~\ref{QBell-on-off}--\ref{fig:hybridparity}.}
  \label{fig:comparison}
\end{figure}

We have demonstrated that an optical hybrid entanglement can be practically used to violate a CHSH Bell inequality with available PNR detectors. In the test, each mode of $\lvert\Psi\rangle$, Eq.~(\ref{HEC}), interferes with a local coherent state $\ket{\alpha}$ and $\ket{\beta}$, respectively, on variable beam splitters of reflectivities $r_a$ and $r_b$, and subsequently measured. While $r_{a,b}$ act as the Bell test settings, splitting the detection outcomes into two sets of zero and non-zero or even/odd numbers of photons allows us to realize binary on/off measurements. This test requires $|\beta|^2>|\gamma|^2$ and small finetuned
beam splitter reflectivities, otherwise the entanglement is decreased due to correlations with other modes caused by interference.
Although the proposed test is challenging to implement due to inevitable losses in the optical paths and imperfect detection, it is within reach of current technology, for example, superconducting-nanowire detectors~\cite{Reddy2020} or transition-edge sensors~\cite{Lita2008}.

For the non-homodyne limit with $\alpha=\beta=2$ our result is depicted in Figs.~\ref{QBell-on-off}(a-b), where significant CHSH Bell inequality violations, up to $S=2.57$, are demonstrated for hybrid entangled state $\lvert\Psi\rangle$ with amplitudes $\gamma<0.81$, and on/off measurements with detection efficiencies above $0.87$. By increasing the value of $\alpha,\beta$ the optimal Bell value also increases.

Our main result is shown in Figs.~\ref{QBell-on-off}(c-d) computed for the same on/off measurement but this time in the homodyne limit, i.e. $r_{a,b}\to0$ and $\alpha,\beta\to\infty$, where the local interferences of $\lvert\Psi\rangle$ with coherent states amount to displacement operations~\cite{Paris1996}. Here we have Bell violations, up to $S=2.71$ in ideal circumstances, which are demonstrated for states with amplitudes $\gamma<1.25$, and on/off measurements with detection efficiencies above $0.82$.
In realistic conditions, for example for $\gamma=0.44$ and efficiencies $\eta_{a,b}>0.95$, it gave the Bell violation of $S=2.51$, which can be obtained with the current experimental setups.


In Fig.~\ref{QBell-parity} we present the result for parity measurement. In the ideal lossless conditions, this allows one to perform the even/odd Bell test for arbitrarily large amplitude $\gamma$, reaching the maximal
value of $S=2.44$. However, this measurement is quickly spoiled by losses that remove photons from the state and so these violations are very fragile and vanish quickly by experimental imperfections.

The results presented in Figs.~\ref{fig:hybridonoff}--\ref{fig:hybridparity} are interesting for studying the theory and properties of hybrid entanglement. We enhanced our two measurement strategies with a general qubit measurement performed in the single-photon subsystem. This allowed us to maximally violate the CHSH inequality, up to $S=2\sqrt{2}$. For example, it lets us decrease the required minimal efficiency of the measurement in the on/off test to $\eta_{a,b}\geq 0.8$ and in the parity test to  $\eta_{a,b}\geq 0.88$.

Our numerical computations are summarised in Fig.~\ref{fig:comparison}, where we show a comparison of the thresholds above which CHSH violation occurs, as a function of both the detection efficiency, as well as the amplitude~$\gamma$. Panel~(a) displays the lines obtained when the local interference is done with a coherent state with finite intensity, while in panel~(b) we show the counterpart in the homodyne limit. In both cases, we find that on/off measurements are most robust to losses, while parity measurements allow violations for higher values of~$\gamma$. Moreover, regardless of the type of measurement, the hybrid strategies provide a clear enhancement of the results, lowering the efficiency required to observe CHSH violations. The Supplemental Material includes a table with all the detailed results of the computations.

Our results give interesting insights from both fundamental and technological standpoints. They show ways to study the quantum nonlocality of optical hybrid entangled states, as well as of more general complex bipartite entanglement merging the CV and DV realms.

\begin{acknowledgments}
M.M., A.B. and M.S. were supported by the European Union’s Horizon 2020 research and innovation programme under the Marie Skłodowska-Curie project ``AppQInfo'' No. 956071. M.S. and A.B. were supported by the National Science Centre ``Sonata Bis'' project No. 2019/34/E/ST2/00273, the QuantERA II Programme that has received funding from the European Union’s Horizon 2020 research and innovation programme under Grant Agreement No 101017733, project ``PhoMemtor'' No. 2021/03/Y/ST2/00177, as well as by the Foundation for Polish Science ``First Team'' project No. POIR.04.04.00-00- 220E/16-00 (originally FIRST TEAM/2016-2/17). J.C.L.C. was supported by the Polish
National Agency for Academic Exchange (NAWA) under project TULIP
with number PPN/ULM/2020/1/00235 and from the Polish National
Science Center (NCN) ``Sonatina'' project CARAMEL with number
2021/40/C/ST2/00155. J.L. is a member of the Institut Universitaire de France. This work was supported by the ANR in the framework of France 2030 (ANR-22-PETQ-0011) and via ShoQC project (19-QUAN-0005-05). We gratefully acknowledge Poland’s high-performance computing infrastructure PLGrid (HPC Centers: ACK Cyfronet AGH) for providing computer facilities and support within computational grant no. PLG/2023/016211. We thank A.~Mikos-Nuszkiewicz for his discussions.
\end{acknowledgments}

\end{document}


\title{Supplementary Material:\\ CHSH Bell Tests For Optical Hybrid Entanglement}

\author{Morteza~Moradi}
\affiliation{Institute of Informatics, Faculty of Mathematics, Informatics and Mechanics, University of Warsaw, Banacha 2, 02-097 Warsaw, Poland}

\author{Juan Camilo {L\'opez Carre\~no}}
\affiliation{Institute of Theoretical Physics, Faculty of Physics, University of Warsaw, Pasteura 5, 02-093 Warsaw, Poland}
\affiliation{Institute of Informatics, Faculty of Mathematics, Informatics and Mechanics, University of Warsaw, Banacha 2, 02-097 Warsaw, Poland}

\author{Adam Buraczewski}
\affiliation{Institute of Informatics, Faculty of Mathematics, Informatics and Mechanics, University of Warsaw, Banacha 2, 02-097 Warsaw, Poland}

\author{Thomas~McDermott}
\affiliation{Institute of Informatics, Faculty of Mathematics, Informatics and Mechanics, University of Warsaw, Banacha 2, 02-097 Warsaw, Poland}

\author{Beate Elisabeth Asenbeck}
\affiliation{Laboratoire Kastler Brossel, Sorbonne Universit\'e, CNRS, ENS-Universit\'e PSL, Coll\`ege de France, 4 Place Jussieu, 75005 Paris, France}

\author{Julien Laurat}
\affiliation{Laboratoire Kastler Brossel, Sorbonne Universit\'e, CNRS, ENS-Universit\'e PSL, Coll\`ege de France, 4 Place Jussieu, 75005 Paris, France}

\author{Magdalena Stobi\'nska}
\email{magdalena.stobinska@gmail.com}
\thanks{Corresponding author}
\affiliation{Institute of Informatics, Faculty of Mathematics, Informatics and Mechanics, University of Warsaw, Banacha 2, 02-097 Warsaw, Poland}

\date{\today}

\maketitle

\section{Derivations of correlation functions $\langle A_i B_j \rangle $}

To have a physical model of an imperfect photodetector with detection efficiency $\eta$, a beam splitter with a transmission coefficient of $\sqrt{\eta}$ is placed before a perfect photodetector
\begin{equation}
    E^{(n)}_\eta=
    \sum_{m=0}^{\infty} \binom{n+m}{n} \eta^n (1-\eta)^m \ket{n+m}\bra{n+m}.
\end{equation}
We can calculate the effective on/off and parity measurement (including the effect of losses) as
\begin{equation}
    \Pi^{\text{on/off}}_{\eta} 
    = E^{(0)}_\eta-\sum_{n=1}^{\infty} E^{(n)}_\eta 
    = \sum_{m=0}^{\infty} (2(1-\eta)^m-1) \ket{m}\bra{m},\qquad
    \Pi^{\text{on/off}}_{\eta} (\delta)
    = D(\delta) \Pi^{\text{on/off}}_{\eta}  D^{\dagger}(\delta),
\end{equation}

\begin{equation}
    \Pi^{\text{parity}}_{\eta} 
    = \sum_{n=0}^{\infty} (-1)^n E^{(n)}_\eta 
    = \sum_{m=0}^{\infty} (1-2  \eta)^m \ket{m}\bra{m},\qquad
    \Pi^{\text{parity}}_{\eta} (\delta)
    = D(\delta) \Pi^{\text{parity}}_{\eta}  D^{\dagger}(\delta).
\end{equation}
For the hybrid measurement case, to eliminate the possibility of the detection loophole, we assign a value of $+1$ for a `no-detection' outcome. So the effective hybrid measurement is
\begin{equation}
    \Pi^{\text{hybrid}}_{\eta} (\theta,\phi) 
    = \eta (\sin{\theta} \cos{\phi} \sigma_X + \sin{\theta} \sin{\phi}\sigma_Y + \cos{\theta} \sigma_Z)
    + (1-\eta) \mathbb{I}.
\end{equation}
Now we can calculate the correlators assuming the homodyne limit
\begin{equation}\label{CorrelatorDef}
    \langle A_i B_j \rangle 
    = \mathrm{Tr}\{\rho (A_i\otimes B_j)\} 
    = \bra{\Psi} A^{\text{eff}}_i  \otimes B^{\text{eff}}_j \ket{\Psi},
\end{equation}
where $A^{\text{eff}}_i$ and $B^{\text{eff}}_j$ are effective measurement projectors including the effect of losses. Inserting $\ket{\Psi} = \frac{1}{\sqrt{2}}\bigl(\ket{0}\ket{\text{cat}^-} + \ket{1}\ket{\text{cat}^+}\bigr)$ in Eq. (\ref{CorrelatorDef}), we get
\begin{align}\label{CorrelatorEffProjectors}
\langle A_i B_j \rangle 
& =\left( +\frac{\bra{0} A^{\text{eff}}_i \ket{0}}{2N^2_-}
+\frac{\bra{0}A^{\text{eff}}_i\ket{1} +\bra{1}A^{\text{eff}}_i\ket{0}}{2N_+N_-}
+\frac{\bra{1}A^{\text{eff}}_i\ket{1}}{2N^2_+}\right)
\bra{+\gamma} B^{\text{eff}}_j \ket{+\gamma}
\nonumber\\
& + \left( +\frac{\bra{0}A^{\text{eff}}_i\ket{0}}{2N^2_-}
-\frac{\bra{0}A^{\text{eff}}_i\ket{1} +\bra{1}A^{\text{eff}}_i\ket{0}}{2N_+N_-}
+\frac{\bra{1}A^{\text{eff}}_i\ket{1}}{2N^2_+}\right)
\bra{-\gamma} B^{\text{eff}}_j \ket{-\gamma}
\nonumber\\
& + \left( -\frac{\bra{0}A^{\text{eff}}_i\ket{0}}{2N^2_-}
+\frac{\bra{0}A^{\text{eff}}_i\ket{1} -\bra{1}A^{\text{eff}}_i\ket{0}}{2N_+N_-}
+\frac{\bra{1}A^{\text{eff}}_i\ket{1}}{2N^2_+}\right)
\bra{+\gamma} B^{\text{eff}}_j \ket{-\gamma}
\nonumber\\
& + \left( -\frac{\bra{0}A^{\text{eff}}_i\ket{0}}{2N^2_-}
-\frac{\bra{0}A^{\text{eff}}_i\ket{1} -\bra{1}A^{\text{eff}}_i\ket{0}}{2N_+N_-}
+\frac{\bra{1}A^{\text{eff}}_i\ket{1}}{2N^2_+}\right)
\bra{-\gamma} B^{\text{eff}}_j \ket{+\gamma}.
\end{align}

\subsection{CHSH test with on/off measurement}

Here the effective measurements are 
$A^{\text{eff}}_i =  \Pi^{\text{on/off}}_{\eta_a} (\delta_{\alpha_i})$, $B^{\text{eff}}_j = \Pi^{\text{on/off}}_{\eta_b} (\delta_{\beta_j})$ for $i,j \in \{1,2\}$.
Assuming $\phi_i$ and $w_j$ are the phase of $\delta_{\alpha_i}$ and $\delta_{\beta_j}$
(i.e.  $\delta_{\alpha_i} = |\delta_{\alpha_i}| e^{i \phi_i}$, $\delta_{\beta_j} = |\delta_{\beta_j}| e^{i w_j}$), a straightforward calculation gives us
\begin{align}\label{OpticalOnOffQPart}
    &\bra{0}\Pi^{\text{on/off}}_{\eta_a} (\delta_{\alpha_i}) \ket{0}
    = 2e^{-\eta_a|\delta_{\alpha_i}|^2}-1,
     \nonumber\\
     &\bra{0}\Pi^{\text{on/off}}_{\eta_a} (\delta_{\alpha_i}) \ket{1}
     =\bra{1}\Pi^{\text{on/off}}_{\eta_a} (\delta_{\alpha_i}) \ket{0}^*
     =2\eta_a \delta_{\alpha_i}^* e^{-\eta_a|\delta_{\alpha_i}|^2},
     \nonumber\\
     &\bra{1}\Pi^{\text{on/off}}_{\eta_a} (\delta_{\alpha_i}) \ket{1}
     = 2(1-\eta_a+\eta^2_a|\delta_{\alpha_i}|^2)e^{-\eta_a|\delta_{\alpha_i}|^2}-1,
\end{align}
for the quantum part, and
\begin{align}\label{OpticalOnOffCPart}
    & \bra{\pm\gamma} \Pi^{\text{on/off}}_{\eta_b} (\delta_{\beta_j}) \ket{\pm\gamma}
    = 2e^{-\eta_b|\pm \gamma - \delta_{\beta_j}|^2}-1,
     \nonumber\\
     &\bra{\pm\gamma} \Pi^{\text{on/off}}_{\eta_b} (\delta_{\beta_j}) \ket{\mp\gamma}
     = 2 e^{-\eta_b |\delta_{\beta_j}|^2-(2-\eta_b)\gamma^2\pm 2 i \eta_b \gamma |\delta_{\beta_j}|\sin{w_j}}
     -e^{-2\gamma^2},
\end{align}
for the macroscopic part. Finally, by replacing (\ref{OpticalOnOffQPart}), (\ref{OpticalOnOffCPart}) in (\ref{CorrelatorEffProjectors}) we can obtain the explicit formula of correlators
\begin{align}\label{AiBjOpticalOnOff}
  &\langle A_i B_j \rangle^{\text{on/off}}_{\text{optical}} =\nonumber\\
  & -2 e^{-\eta_a |\delta_{\alpha_i}|^2 -\eta_b (|\delta_{\beta_j}|^2+\gamma^2)}
  \cosh(2 \eta_b \gamma |\delta_{\beta_j}| \cos w_j)\hspace{1mm}
  \frac{\eta_a(1-\eta_a|\delta_{\alpha_i}|^2)(1-e^{-2\gamma^2})-2+e^{\eta_a |\delta_{\alpha_i}|^2}}{1-e^{-4\gamma^2}}
  \nonumber\\
  & + 2e^{-\eta_a |\delta_{\alpha_i}|^2 -\eta_b(|\delta_{\beta_j}|^2+\gamma^2)}
  \sinh(2 \eta_b \gamma |\delta_{\beta_j}| \cos w_j) \hspace{1mm}
  \frac{2\eta_a |\delta_{\alpha_i}| \cos\phi_i}{\sqrt{1-e^{-4\gamma^2}}}
  \nonumber\\
  & - 2e^{-\eta_a |\delta_{\alpha_i}|^2 -\eta_b|\delta_{\beta_j}|^2-(2-\eta_b)\gamma^2}
  \cos(2 \eta_b \gamma |\delta_{\beta_j}|\sin w_j)
  \frac{\eta_a(1-\eta_a|\delta_{\alpha_i}|^2)(1-e^{-2\gamma^2})+2e^{-2\gamma^2}-e^{\eta_a |\delta_{\alpha_i}|^2-2\gamma^2}}{1-e^{-4\gamma^2}}
  \nonumber\\
  & + 2e^{-\eta_a |\delta_{\alpha_i}|^2 -\eta_b|\delta_{\beta_j}|^2-(2-\eta_b)\gamma^2}
 \sin(2 \eta_b \gamma |\delta_{\beta_j}|\sin w_j)
  \frac{2\eta_a |\delta_{\alpha_i}|\sin\phi_i}{\sqrt{1-e^{-4\gamma^2}}} + 1 -e^{-\eta_a |\delta_{\alpha_i}|^2 }(2-\eta_a(1-\eta_a |\delta_{\alpha_i}|^2 )).
\end{align}

\subsection{CHSH test with parity measurement}

Here the effective measurements are 
$A^{\text{eff}}_i =  \Pi^{\text{parity}}_{\eta_a} (\delta_{\alpha_i})$, $B^{\text{eff}}_j = \Pi^{\text{parity}}_{\eta_b} (\delta_{\beta_j})$ for $i,j \in \{1,2\}$
\begin{align}\label{OpticalParityQPart}
    &\bra{0}\Pi^{\text{parity}}_{\eta_a} (\delta_{\alpha_i}) \ket{0}
    = e^{-2\eta_a|\delta_{\alpha_i}|^2},
     \nonumber\\
     &\bra{0}\Pi^{\text{parity}}_{\eta_a} (\delta_{\alpha_i}) \ket{1}
     =\bra{1}\Pi^{\text{parity}}_{\eta_a} (\delta_{\alpha_i}) \ket{0}^*
     =2\eta_a \delta_{\alpha_i}^* e^{-2\eta_a|\delta_{\alpha_i}|^2},
     \nonumber\\
     &\bra{1}\Pi^{\text{parity}}_{\eta_a} (\delta_{\alpha_i}) \ket{1}
     = (1-2\eta_a+4\eta^2_a|\delta_{\alpha_i}|^2)e^{-2\eta_a|\delta_{\alpha_i}|^2},
\end{align}
for the quantum part, and
\begin{align}\label{OpticalParityCPart}
    & \bra{\pm\gamma} \Pi^{\text{parity}}_{\eta_b} (\delta_{\beta_j}) \ket{\pm\gamma}
    = e^{-2\eta_b|\pm \gamma - \delta_{\beta_j}|^2},
     \nonumber\\
     &\bra{\pm\gamma} \Pi^{\text{parity}}_{\eta_b} (\delta_{\beta_j}) \ket{\mp\gamma}
     = e^{-2\eta_b |\delta_{\beta_j}|^2-2(1-\eta_b)\gamma^2\pm 4 i \eta_b \gamma |\delta_{\beta_j}|\sin{w_j}},
\end{align}
for the classical part. By replacing (\ref{OpticalParityQPart}), (\ref{OpticalParityCPart}) in (\ref{CorrelatorEffProjectors}) we can obtain the explicit form of correlators
\begin{align}\label{AiBjOpticalParity}
  &\langle A_i B_j \rangle^{\text{parity}}_{\text{optical}} =\nonumber\\
  & - e^{-2\eta_a |\delta_{\alpha_i}|^2 -2\eta_b (|\delta_{\beta_j}|^2+\gamma^2)}
  \cosh(4 \eta_b \gamma |\delta_{\beta_j}| \cos w_j)\hspace{1mm}
  \frac{\eta_a(1-2\eta_a|\delta_{\alpha_i}|^2)(1-e^{-2\gamma^2})-1}{1-e^{-4\gamma^2}}
  \nonumber\\
  & + e^{-2\eta_a |\delta_{\alpha_i}|^2 -2\eta_b(|\delta_{\beta_j}|^2+\gamma^2)}
 \sinh(4 \eta_b \gamma |\delta_{\beta_j}| \cos w_j) \hspace{1mm}
  \frac{2\eta_a |\delta_{\alpha_i}| \cos\phi_i}{\sqrt{1-e^{-4\gamma^2}}}
  \nonumber\\
  & - e^{-2\eta_a |\delta_{\alpha_i}|^2 -2\eta_b|\delta_{\beta_j}|^2-2(1-\eta_b)\gamma^2}
  \cos(4 \eta_b \gamma |\delta_{\beta_j}|\sin w_j)
  \frac{\eta_a(1-2\eta_a|\delta_{\alpha_i}|^2)(1-e^{-2\gamma^2})+e^{-2\gamma^2}}{1-e^{-4\gamma^2}}
  \nonumber\\
  & + e^{-2\eta_a |\delta_{\alpha_i}|^2 -2\eta_b|\delta_{\beta_j}|^2-2(1-\eta_b)\gamma^2}
 \sin(4 \eta_b \gamma |\delta_{\beta_j}|\sin w_j)
  \frac{2\eta_a |\delta_{\alpha_i}|\sin\phi_i}{\sqrt{1-e^{-4\gamma^2}}}.
\end{align}
Optimized value of fully optical CHSH test using Eqs.~(\ref{AiBjOpticalOnOff}) and (\ref{AiBjOpticalParity}) for different values of gamma and losses are shown in Figs.~\ref{fig:optical_onoff}--\ref{fig:optical_parity}.

\subsection{Hybrid CHSH test with on/off measurements}

Here we consider hybrid measurement in the quantum part and on/off measurement in the macroscopic part. So the corresponding effective measurements are $A^{\text{eff}}_i = \Pi^{\text{hybrid}}_{\eta} (\theta_i,\phi_i) $ and $B^{\text{eff}}_j = \Pi^{\text{on/off}}_{\eta_b} (\delta_{\beta_j})$ for $i,j \in \{1,2\}$
\begin{align}\label{HybridOnOffQPart}
    &\bra{0} A^{\text{eff}}_i \ket{0}
    = (1-\eta_a) + \eta_a \cos{\theta_i},
     \nonumber\\
     &\bra{0} A^{\text{eff}}_i \ket{1}
     =\bra{1} A^{\text{eff}}_i \ket{0}^*
     = \eta_a e^{-i\phi_i} \sin{\theta_i},
     \nonumber\\
     &\bra{1} A^{\text{eff}}_i \ket{1}
     =  (1-\eta_a) - \eta_a \cos{\theta_i}.
\end{align}
Putting (\ref{HybridOnOffQPart}), (\ref{OpticalOnOffCPart}) in (\ref{CorrelatorEffProjectors}) we have
\begin{align}\label{AiBjHybridOnOff}
  &\langle A_i B_j \rangle^{\text{on/off}}_{\text{hybrid}} =\nonumber\\
  & +2 e^{-\eta_b (|\delta_{\beta_j}|^2+\gamma^2)}
  \cosh(2 \eta_b \gamma |\delta_{\beta_j}| \cos w_j)\hspace{1mm}
  \frac{(1-\eta_a)+\eta_a \cos\theta_i e^{-2\gamma^2}}{1-e^{-4\gamma^2}}
  \nonumber\\
  & - 2e^{-\eta_b(|\delta_{\beta_j}|^2+\gamma^2)}
 \sinh(2 \eta_b \gamma |\delta_{\beta_j}| \cos w_j) \hspace{1mm}
  \frac{\eta_a\sin\theta_i \cos\phi_i}{\sqrt{1-e^{-4\gamma^2}}}
  \nonumber\\
  & - 2e^{-\eta_b|\delta_{\beta_j}|^2-(2-\eta_b)\gamma^2}
  \cos(2 \eta_b \gamma |\delta_{\beta_j}|\sin w_j)
  \frac{(1-\eta_a)e^{-2\gamma^2}+\eta_a \cos\theta_i}{1-e^{-4\gamma^2}}
  \nonumber\\
  & + 2e^{-\eta_b|\delta_{\beta_j}|^2-(2-\eta_b)\gamma^2}
 \sin(2 \eta_b \gamma |\delta_{\beta_j}|\sin w_j)
  \frac{\eta_a\sin\theta_i\sin\phi_i}{\sqrt{1-e^{-4\gamma^2}}} + \eta_a-1.
\end{align}

\subsection{Hybrid CHSH test with parity measurements}

Here the effective measurements are $A^{\text{eff}}_i =\Pi^{\text{hybrid}}_{\eta} (\theta_i,\phi_i) $ and $B^{\text{eff}}_j = \Pi^{\text{parity}}_{\eta_b} (\delta_{\beta_j})$ for $i,j \in \{1,2\}$. To calculate correlators, we should replace (\ref{HybridOnOffQPart}), (\ref{OpticalParityCPart}) in (\ref{CorrelatorEffProjectors})
\begin{align}\label{AiBjHybridParity}
  &\langle A_i B_j \rangle^{\text{parity}}_{\text{hybrid}} =\nonumber\\
  & + e^{-2\eta_b (|\delta_{\beta_j}|^2+\gamma^2)}
  \cosh(4 \eta_b \gamma |\delta_{\beta_j}| \cos w_j)\hspace{1mm}
  \frac{(1-\eta_a)+\eta_a \cos\theta_i e^{-2\gamma^2}}{1-e^{-4\gamma^2}}
  \nonumber\\
  & - e^{-2\eta_b(|\delta_{\beta_j}|^2+\gamma^2)}
 \sinh(4 \eta_b \gamma |\delta_{\beta_j}| \cos w_j) \hspace{1mm}
  \frac{\eta_a\sin\theta_i \cos\phi_i}{\sqrt{1-e^{-4\gamma^2}}}
  \nonumber\\
  & - e^{-2\eta_b|\delta_{\beta_j}|^2-2(1-\eta_b)\gamma^2}
  \cos(4 \eta_b \gamma |\delta_{\beta_j}|\sin w_j)
  \frac{(1-\eta_a)e^{-2\gamma^2}+\eta_a \cos\theta_i}{1-e^{-4\gamma^2}}
  \nonumber\\
  & + e^{-2\eta_b|\delta_{\beta_j}|^2-2(1-\eta_b)\gamma^2}
 \sin(4 \eta_b \gamma |\delta_{\beta_j}|\sin w_j)
  \frac{\eta_a\sin\theta_i\sin\phi_i}{\sqrt{1-e^{-4\gamma^2}}}.
\end{align}
Optimized value of hybrid CHSH test using Eq.(\ref{AiBjHybridOnOff}),(\ref{AiBjHybridParity}) for different values of gamma and losses are shown in Figs.~\ref{fig:hybrid_onoff}--\ref{fig:hybrid_parity}. Finally, we provide a comparison of all of these 4 cases in the table \ref{Comparision table}.

\begin{table}[]
    \centering
    \begin{tabular}{|c|c|c|c|c|c|c|}
  \hline
   Alice's & Bob's & Some features & & Optimal $\gamma$ & Max & optimal $\gamma$ \\
   & & of optimal &  $S>2$  & corresponds & S & corresponds \\
   POVM &  POVM & setting &  & to Min $\eta$ & &  to Max S  \\
  \hline
    & & & $\eta_a=\eta_b\ge 81.9\% $ & 0.60 & & \\
    \cline{4-5}
    & & $\phi_1=0,\phi_2=\pi$ &  $\eta_a = 100\% , \hspace{1mm} \eta_b\ge 63\% $ & 0.67 & 2.71 &  0.40 \\
    \cline{4-5}
    $\Pi^{\text{on/off}}_{\eta_a} (\delta_{\alpha_i})$ & $\Pi^{\text{on/off}}_{\eta_b} (\delta_{\beta_j})$ & $w_1=0,w_2=\pi$ & $\eta_a\ge 66.5\% , \hspace{1mm} \eta_b = 100\%$ & 0.54 & & \\
    \cline{4-7}
    & &  $ \delta_{\alpha_i} , \delta_{\beta_j}, \gamma \in \mathbb{R}$ & $\eta_a = 90\% , \hspace{1mm} \eta_b \ge 73.1\% $ & 0.63 & 2.50 & 0.42 \\
    \cline{4-7}
    & &  & $\eta_a \ge 74.8\% , \hspace{1mm} \eta_b = 90\% $ & 0.58 & 2.52 & 0.46 \\
    \cline{4-7}
  \hline
     & & & $\eta_a=\eta_b\ge 94\% $ & 0.54 &  & \\
    \cline{4-5}
    & & $\phi_1=-\phi_2=\frac{\pi}{2}$ &  $\eta_a = 100\% , \hspace{1mm} \eta_b\ge 88.4\% $ & 0.35 & 2.44 & $+\infty$ \\
    \cline{4-5}
    $\Pi^{\text{parity}}_{\eta_a} (\delta_{\alpha_i})$ & $\Pi^{\text{parity}}_{\eta_b} (\delta_{\beta_j})$ & $w_1=-w_2=\frac{\pi}{2}$ & $\eta_a\ge 78.7\% , \hspace{1mm} \eta_b = 100\%$ & $+\infty$ & &   \\
    \cline{4-7}
    & &  $ i \delta_{\alpha_i} , i\delta_{\beta_j}, \gamma \in \mathbb{R}$ & $\eta_a = 95\% , \hspace{1mm} \eta_b \ge 93\% $ & 0.50 & 2.34 &  $+\infty$ \\
    \cline{4-7}
    & &  & $\eta_a \ge 92.6\% , \hspace{1mm} \eta_b = 95\% $ & 0.60 & 2.15 & 0.57  \\
    \cline{4-7}
  \hline
  \hline
    & & & $\eta_a=\eta_b\ge 79.8\% $ & 0.53 & & \\
    \cline{4-5}
    & & $\phi_1=0,\phi_2=\pi$ &  $\eta_a = 100\% , \hspace{1mm} \eta_b > 50\% $ & 1.75 & $2\sqrt{2}$ &  0.416 \\
    \cline{4-5}
    $\Pi^{\text{hybrid}}_{\eta_a} (\theta_i,\phi_i)$ & $\Pi^{\text{on/off}}_{\eta_b} (\delta_{\beta_j})$ & $w_1=0,w_2=\pi$ & $\eta_a > \sqrt{2}/2 , \hspace{1mm} \eta_b = 100\%$ & 0.416 & & \\
    \cline{4-7}
    & &  $ \delta_{\beta_j}, \gamma \in \mathbb{R}$ & $\eta_a = 90\% , \hspace{1mm} \eta_b \ge 65\% $ & 0.69 & 2.55 & 0.416 \\
    \cline{4-7}
    & &  & $\eta_a \ge 74.7\% , \hspace{1mm} \eta_b = 90\% $ & 0.46 & 2.63 & 0.49 \\
    \cline{4-7}
  \hline
     & & & $\eta_a=\eta_b\ge 88.1\% $ & 0.47 &  & \\
    \cline{4-5}
    & & $\phi_1=-\phi_2=\frac{\pi}{2}$ &  $\eta_a = 100\% , \hspace{1mm} \eta_b\ge 78.5\% $ & 0.31 & $2\sqrt{2}$ & $+\infty$ \\
    \cline{4-5}
    $\Pi^{\text{hybrid}}_{\eta_a} (\theta_i,\phi_i)$ & $\Pi^{\text{parity}}_{\eta_b} (\delta_{\beta_j})$ & $w_1=-w_2=\frac{\pi}{2}$ & $\eta_a > \sqrt{2}/2 , \hspace{1mm} \eta_b = 100\%$ & $+\infty$ & &   \\
    \cline{4-7}
    & &  $ i\delta_{\beta_j}, \gamma \in \mathbb{R}$ & $\eta_a = 95\% , \hspace{1mm} \eta_b \ge 82.1\% $ & 0.37 & 2.69 &  $+\infty$ \\
    \cline{4-7}
    & &  & $\eta_a \ge 81.2\% , \hspace{1mm} \eta_b = 95\% $ & 0.67 & 2.42 & 0.77  \\
    \cline{4-7}
  \hline
\end{tabular}
    \caption{Comparison Table in the Homodyne limit}
    \label{Comparision table}
\end{table}

\begin{figure}[p]
    \centering
    \raisebox{4cm}{(a)}
    \includegraphics[width=8cm]{figs/Optical OnOff CHSH test for etaA=etaB.pdf}
    \\
    \raisebox{4cm}{(b)}
    \includegraphics[width=8cm]{figs/Optical OnOff CHSH test for etaA=1.pdf}
    \raisebox{4cm}{(c)}
    \includegraphics[width=8cm]{figs/Optical OnOff CHSH test for etaB=1.pdf}
    \\
    \raisebox{4cm}{(d)}
    \includegraphics[width=8cm]{figs/Optical OnOff CHSH test for etaA=0.90.pdf}
    \raisebox{4cm}{(e)}
    \includegraphics[width=8cm]{figs/Optical OnOff CHSH test for etaB=0.90.pdf}
    \caption{Maximum value of violation in fully optical CHSH Bell test of the hybrid entangled state $\ket{\Psi}$ as a function of amplitude $\gamma$ and total losses $\eta_{a,b}$ in both arms for the on/off measurement and (a) equal losses, $\eta_a=\eta_b$, (b) no losses in mode $a$, $\eta_a=1$, (c) no losses in mode $b$, $\eta_b=1$, (d) 10\% of losses in mode $a$, $\eta_a=0.90$, and (e) 10\% of losses in mode $b$, $\eta_b=0.90$.}
    \label{fig:optical_onoff}
\end{figure}

\begin{figure}[p]
    \centering
    \raisebox{4cm}{(a)}
    \includegraphics[width=7cm]{figs/Optical Parity CHSH test for etaA=etaB.pdf}
    \\
    \raisebox{4cm}{(b)}
    \includegraphics[width=7cm]{figs/Optical Parity CHSH test for etaA=1.pdf}
    \raisebox{4cm}{(c)}
    \includegraphics[width=7cm]{figs/Optical Parity CHSH test for etaB=1.pdf}
    \\
    \raisebox{4cm}{(d)}
    \includegraphics[width=7cm]{figs/Optical Parity CHSH test for etaA=0.95.pdf}
    \raisebox{4cm}{(e)}
    \includegraphics[width=7cm]{figs/Optical parity CHSH test for etaB=0.95.pdf}
    \caption{Maximum value of violation in fully optical CHSH Bell test of the hybrid entangled state $\ket{\Psi}$ as a function of amplitude $\gamma$ and total losses $\eta_{a,b}$ in both arms for the parity measurement and (a) equal losses, $\eta_a=\eta_b$, (b) no losses in mode $a$, $\eta_a=1$, (c) no losses in mode $b$, $\eta_b=1$, (d) 5\% of losses in mode $a$, $\eta_a=0.95$, and (e) 5\% of losses in mode $b$, $\eta_b=0.95$.}
    \label{fig:optical_parity}
\end{figure}

\begin{figure}[p]
    \centering
    \raisebox{4cm}{(a)}
    \includegraphics[width=7cm]{figs/Hybrid OnOff CHSH test for etaA=etaB.pdf}
    \\
    \raisebox{4cm}{(b)}
    \includegraphics[width=7cm]{figs/Hybrid OnOff CHSH test for etaA=1.pdf}
    \raisebox{4cm}{(c)}
    \includegraphics[width=7cm]{figs/Hybrid OnOff CHSH test for etaB=1.pdf}
    \\
    \raisebox{4cm}{(d)}
    \includegraphics[width=7cm]{figs/Hybrid OnOff CHSH test for etaA=0.90.pdf}
    \raisebox{4cm}{(e)}
    \includegraphics[width=7cm]{figs/Hybrid OnOff CHSH test for etaB=0.90.pdf}
    \caption{Maximum value of violation in fully optical CHSH Bell test of the hybrid entangled state $\ket{\Psi}$ as a function of amplitude $\gamma$ and total losses $\eta_{a,b}$ in both arms for the hybrid on/off measurement and (a) equal losses, $\eta_a=\eta_b$, (b) no losses in mode $a$, $\eta_a=1$, (c) no losses in mode $b$, $\eta_b=1$, (d) 10\% of losses in mode $a$, $\eta_a=0.90$, and (e) 10\% of losses in mode $b$, $\eta_b=0.90$.}
    \label{fig:hybrid_onoff}
\end{figure}

\begin{figure}[p]
    \centering
    \raisebox{4cm}{(a)}
    \includegraphics[width=7cm]{figs/Hybrid Parity CHSH test for etaA=etaB.pdf}
    \\
    \raisebox{4cm}{(b)}
    \includegraphics[width=7cm]{figs/Hybrid Parity CHSH test for etaA=1.pdf}
    \raisebox{4cm}{(c)}
    \includegraphics[width=7cm]{figs/Hybrid Parity CHSH test for etaB=1.pdf}
    \\
    \raisebox{4cm}{(d)}
    \includegraphics[width=7cm]{figs/Hybrid Parity CHSH test for etaA=0.95.pdf}
    \raisebox{4cm}{(e)}
    \includegraphics[width=7cm]{figs/Hybrid parity CHSH test for etaB=0.95.pdf}
    \caption{Maximum value of violation in fully optical CHSH Bell test of the hybrid entangled state $\ket{\Psi}$ as a function of amplitude $\gamma$ and total losses $\eta_{a,b}$ in both arms for the hybrid parity measurement and (a) equal losses, $\eta_a=\eta_b$, (b) no losses in mode $a$, $\eta_a=1$, (c) no losses in mode $b$, $\eta_b=1$, (d) 5\% of losses in mode $a$, $\eta_a=0.95$, and (e) 5\% of losses in mode $b$, $\eta_b=0.95$.}
    \label{fig:hybrid_parity}
\end{figure}

\clearpage

\section{Numerical methods}

\subsection{Purpose}

Optical hybrid photon-number entanglement $\ket{\Psi}$, defined by Eqs.~(1)--(2) in the main
text, is a two-mode quantum entanglement, consisting of a discrete-variable qubit (mode $A$), encoded in photon-number states $\ket{0}$ and $\ket{1}$ -- the vacuum and single-photon Fock state, respectively, and mode $B$ that carries two mutually orthogonal continuous-variable states, $\ket{\text{cat}^-}$ and $\ket{\text{cat}^+}$. Both $\ket{\text{cat}^{\pm}}$ are superpositions of coherent states of amplitude $\gamma$, however, with opposite phases. Modes $A$
and $B$ are interfered with coherent beams $\ket{\alpha}$ and
$\ket{\beta}$ on variable beam splitters of reflectivities $r_a$ and
$r_b$, respectively, and measured, Fig~1 in the main text. Our goal is to test whether this state violates
the CHSH Bell inequality $S\leq 2$, Eq.~(3) in the main text, for
several cases:
\begin{enumerate}
\item measurements at macroscopic mode $B$: a) zero/non-zero or b)
  odd/even number of photons registered,
\item measurements at microscopic mode $A$: a) the same as for mode
  $B$ or b) with general qubit measurements applied (cf.~Eq.~(7) of
  the main text),
\item interference on the beam splitters: a) with $0<r_{a,b}<1$ and
  coherent pulses of weak amplitudes
  $\lvert\alpha\rvert^2,\lvert\beta\rvert^2<10\lvert\gamma\rvert^2$
  or b) in the limit of $r_{a,b}\to 0$ and
  $\lvert\alpha\rvert^2,\lvert\beta\rvert^2\gg \lvert\gamma\rvert^2$
  where homodyne limit is achieved.
\end{enumerate}
Additionally, the numerical program takes into account imperfections
in the system, inefficient detection and losses in both modes $A$
and $B$. This allows us to find the dependence of the CHSH value $S$
for realistic conditions and check the minimal requirements for the
experimental system to display a violation. The losses are modeled
collectively, by additional beam splitters of transmitivity $\eta_a$
and $\eta_b$ located in optical paths of the modes $A$ and $B$,
respectively.

\subsection{Data types and algorithms}

The initial stage of the experiment can be written as a quantum
state that contains our hybrid entangled state $\ket{\Psi}$ together with the coherent states that will be used
to take into account the interference and losses of both arms. As
such, the resulting quantum state can be written as
\begin{equation}
  \label{eq:Tue25Apr2023113133CEST}
  \ket{\Psi}_i = \frac{1}{\sqrt{2}} \left(\ket{0, \alpha_1, \alpha_2,
      \mathrm{cat}^-,\beta_1, \beta_2 } + 
    \ket{1, \alpha_1, \alpha_2, \mathrm{cat}^+, \beta_1, \beta_2}
  \right), 
\end{equation}
where the amplitudes~$\alpha_i$ and~$\beta_i$ are the amplitudes of
the coherent beams that interfere with modes~$A$ and~$B$,
respectively. The inclusion of two general beams on each mode
allowed us to verify take into account losses either before the
interference with a coherent beam (i.e.,
letting~$\alpha_1 = \beta_1=0$ and $\alpha_2, \beta_2\neq 0$), or
before the detection (i.e., letting~$\alpha_1,\beta_1\neq0$ and
$\alpha_2= \beta_2= 0$). Note that the notation used in
Eq.~(\ref{eq:Tue25Apr2023113133CEST}) takes implicitly the tensor
products among the various modes, namely
\begin{equation}
  \label{eq:Tue25Apr2023113924CEST}
  \ket{0, \alpha_1, \alpha_2,
    \mathrm{cat}^-,\beta_1, \beta_2} \equiv \ket{0}_A \otimes
  \ket{\alpha_1}_2 \otimes \ket{\alpha_2}_3 \otimes
  \ket{\mathrm{cat}^-}_B \otimes \ket{\beta_1}_5 \otimes
  \ket{\beta_2}_6.
\end{equation}
The beam splitters placed along the optical paths are associated to
unitary transformations involving two modes at the time. Thus, we
can define two pairs of operators
\begin{subequations}
  \begin{align}
    \mathcal{U}_1(\varphi_1) = \exp \left[i\varphi_1 \left( \ud{a_A}
    a_2 + \ud{a_2}a_A\right) \right], \quad \quad &  \quad \quad
    \mathcal{U}_2(\varphi_2) = \exp \left[i\varphi_2 \left( \ud{a_A}
    a_3 + \ud{a_3}a_A\right) \right], \\
    \mathcal{U}_3(\varphi_3) = \exp \left[i\varphi_3 \left( \ud{a_B}
    a_5 + \ud{a_5}a_B\right) \right], \quad \quad &  \quad \quad
    \mathcal{U}_4(\varphi_4) = \exp \left[i\varphi_4 \left( \ud{a_B}
    a_6 + \ud{a_6}a_B\right) \right],
  \end{align}
\end{subequations}
where~$\ud{a_c}$ and~$a_c$ are the creation and annihilation
operators associated to mode~$c$ with~$c \in \{ A,B,2,3,5,6 \}$,
and the parameters~$\varphi_j$ are related to the reflection
coefficients of the beam splitters through the
relation~$r_j = \sin^2(\varphi_j/2)$. Thus, using the similarity
relations between unitary operators and displacement operators we
find, for instance
\begin{subequations}
  \label{eq:Tue25Apr2023121322CEST}
  \begin{align}
  \mathcal{U}_3 \ket{\mathrm{cat}^\pm,\beta_1} &\equiv
  \frac{1}{N_\pm} \left (\mathcal{U}_3
    \ket{\gamma,\beta_1} \pm \mathcal{U}_3\ket{-\gamma,\beta_1}\right),\\
    & = \frac{1}{N_\pm} \left (\ket{\gamma \sqrt{t_3} + i
      \beta_1 
      \sqrt{r_3}, \beta_1 \sqrt{t_3} + i \gamma \sqrt{r_3}} \pm
      \ket{-\gamma \sqrt{t_3} + i \beta_1 \sqrt{r_3}, \beta_1
      \sqrt{t_3} - i \gamma \sqrt{r_3}} 
      \right),
  \end{align}
\end{subequations}
where $N_\pm = \sqrt{2(1\pm e^{-2|\gamma|^2})}$ is the
normalization constant of the cat states, and we have the
notation~$t_k = 1- r_k$ to indicate the transmission coefficient of
the beam splitter. Note that each term of the superposition of the
resulting state in Eq.~(\ref{eq:Tue25Apr2023121322CEST}) consists of
the tensor produce of two coherent states.  Moreover, the
transformations for the part of the microscopic mode yield, for
example
\begin{subequations}
  \begin{align}
    \mathcal{U}_1 \ket{1, \alpha_1} & \equiv \mathcal{U}_1 \ud{a_A}
    \ket{0, \alpha_1},\\
    & = \mathcal{U}_1 \ud{a_A} \ud{\mathcal{U}_1} \mathcal{U}_1
      \ket{0, \alpha_1},\\
    & = (\ud{a_A}\sqrt{t_1} + i \ud{a_2}\sqrt{r_1}) \ket{i\alpha_1
      \sqrt{r1}, \alpha_1\sqrt{t_1}},
  \end{align}
\end{subequations}
where we find that the resulting state is a superposition of
creation operators acting on a pair of coherent states.

The quantum state of the experiment after the two modes have been
interfered with their respective coherent fields is then given by
\begin{equation}
  \label{eq:Tue25Apr2023123653CEST}
  \ket{\Psi}_f = \mathcal{U}_2 \mathcal{U}_1 \mathcal{U}_4
  \mathcal{U}_3 \ket{\Psi}_i,
\end{equation}
where~$\ket{\Psi}_i$ is the quantum state given in
Eq.~(\ref{eq:Tue25Apr2023113133CEST}) and, for simplicity, we have
omitted the dependence on the reflectivity coefficients on the
unitary operators~$\mathcal{U}_k$. Finally, we performed the partial
trace over the modes~$2, 3, 5$ and~$6$ to obtain the density matrix
of interest, namely
\begin{equation}
  \label{eq:Tue25Apr2023124306CEST}
  \rho_{A,B} = \Tr_{2,3,5,6} \left \lbrace
    \ket{\Psi}_f\bra{\Psi}_f\right \rbrace,
\end{equation}
which is a mixed state of the modes~$A$ and~$B$, and we have used
the notation~$\Tr_k$ to indicate a trace over mode~$k$. Note that the
order in which the traces are taken does not change the final
result. Using the quantum state in
Eq.~(\ref{eq:Tue25Apr2023124306CEST}) we can evaluate the mean
values as
\begin{equation}
  \label{eq:Tue25Apr2023124742CEST}
  \mean{A_iB_j}^\mathrm{on/off} = \Tr_{A,B} \left \lbrace \rho_{A,B}
    \Pi_{A,i}^\mathrm{on/off}\otimes \Pi_{B,j}^\mathrm{on/off} \right
  \rbrace 
  \quad \quad \mathrm{and} \quad \quad
  \mean{A_iB_j}^\mathrm{parity} = \Tr_{A,B} \left \lbrace \rho_{A,B}
    \Pi_{A,i}^\mathrm{parity}\otimes \Pi_{B,j}^\mathrm{parity} \right
  \rbrace,
\end{equation}
where the operators~$\Pi_{A,i}^\mathrm{on/off}$ and
$\Pi_{A,i}^\mathrm{parity}$ are as defined in the previous section.

Finally, we prepared numerical routines which computed~$S$ in terms
of the mean values given in Eq.~(\ref{eq:Tue25Apr2023124742CEST}),
and a function of the parameters $\gamma$, $\alpha$, $\beta$,
$\eta_{a,b}$, $r_{a,b}$ for the given measurement strategy. Note
that~$\eta_{a,b}$ refer to the transmission coefficients associated
to the beam splitters used to take into account losses on each path,
whereas~$r_{a,b}$ correspond to the reflection coefficients of the
beam splitters used to make the coherent state interfere with the
signal.

To find the optimal parameters for achieving maximal CHSH violation,
we used the Nelder--Mead method, also known as `downhill simplex'
approach, to optimize $S$ with respect to parameters
$(r_{a_1},r_{a_2},r_{b_1},r_{b_2})$ for given $\gamma$, $\alpha$,
$\beta$ and $\eta_{a,b}$ in the regime of low amplitudes of coherent
beams, and parameters
$(\delta_{\alpha_1},\delta_{\alpha_2},\delta_{\beta_1},\delta_{\beta_2})$
in the homodyne limit, respectively. Since this algorithm
requires an initial point, we used a fixed value of these parameters
resulting from the optimization of the first point (typically, for
large~$\eta$ and small~$\gamma$), and then the previously computed
value in subsequent computations for neighbouring parameters.

\subsection{Program structure}

The numerical program was written in Python 3 programming language
using imperative structural programming method with utilization of
\verb|NumPy| and \verb|SciPy| packages. To obtain the result, it
explicitly realizes the following steps.
\begin{enumerate}
\item For the given set of input arguments, such as the measurement
  used (on/off or parity), amplitudes of local oscillators $\alpha$
  and $\beta$ and ranges of the classical state amplitude
  $[\gamma_\text{min},\gamma_\text{max}]$ as well as efficiencies
  $[{\eta_b}_\text{min},{\eta_b}_\text{max}]$ and $\eta_a$, it
  iterates over $\gamma$ and $\eta_b$ with a given step
  $\Delta_\gamma$ and $\Delta_{\eta_b}$.
\item For each value of
  $\gamma=\gamma_\text{min}+\Delta_\gamma\cdot x$ and
  $\eta_b={\eta_b}_\text{min}+\Delta_{\eta_b}\cdot y$, where
  $x,y=0,1,2,\dots$, it calls the Nelder--Mead optimizer built into
  the \verb|SciPy| library, passing one of the routines selecting
  the measurement strategy to it.
\item The routine called by the optimizer computes the mean values,
  as given in Eq.~(\ref{eq:Tue25Apr2023124742CEST}), and
  evaluates~$S$ as in Eq.~(3) of the main text.
\item After finding the optimal value of $S$, the program saved the
  result along with the parameters used in comma-separated-value
  (CSV) files.
\end{enumerate}
Additionally, the program measured its runtime and provides it along
the output data files.

\subsection{Running computations and data postprocessing}

The program was run in batches for a range of input arguments on a
supercomputing cluster at the ACK Cyfronet AGH computing center. For
each value of parameters, a separate, single-threaded process was
used. For scheduling we used \verb|SLURM| queuing system. The files
were then transferred to a laptop and the final plots were prepared either 
in Wolfram Mathematica or Matplotlib, and annotated in Inkscape.

Typical time required to run a single computation was around 2
seconds per point optimized.

\section{Numerical results}

\begin{figure}
    \centering
    \includegraphics[width=8cm]{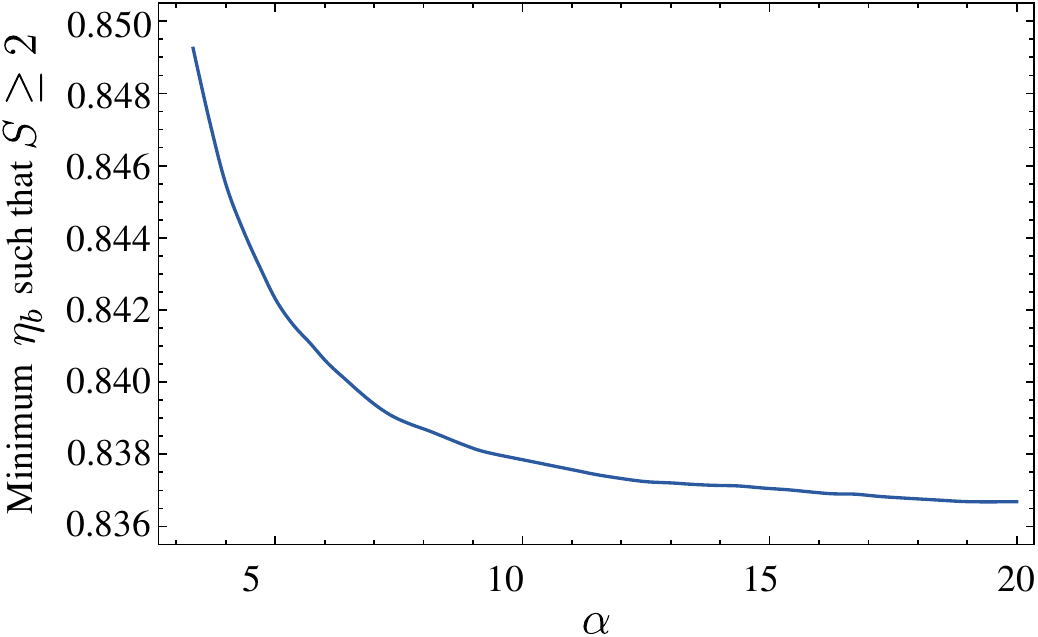}
    \caption{Dependency of the minimal efficiency $\eta_b$ on coherent beam amplitude $\alpha$ for which CHSH Bell inequality violation is observed.}
    \label{fig:minimal_efficiency}
\end{figure}

The results of the computations of the optimized value of the CHSH parameter $S$ for on/off and parity measurements, several values of local oscillator amplitudes $\alpha$, $\beta$ and losses $\eta_{a,b}$ along with the optimal settings $(r_{a_1},r_{a_2},r_{b_1},r_{b_2})$ found by our numerical program are gathered in Tab.~\ref{tab:normal_regime}. Values computed for the displacement limit $r\to 0$, $\alpha,\beta\to\infty$ together with the optimal displacement settings $(\delta_{\alpha_1},\delta_{\alpha_2},\delta_{\beta_1},\delta_{\beta_2})$ are presented in Tab.~\ref{tab:displacement_regime}. Finally, Tab.~\ref{tab:hybrid_normal_regime} and Tab.~\ref{tab:hybrid_displacement_regime} show the optimal settings for the hybrid scheme, in which the microscopic subsystem is probed with general qubit measurements and the macroscopic part -- with on/off and parity measurements, for moderate values of $\alpha,\beta$ and in the displacement limit, respectively.
Fig.~\ref{fig:minimal_efficiency} also depicts the relationship between the minimum requirements for the detection efficiency and the amplitude of coherent beams used for interference which allows one to find the requirements and optimal parameters of the experimental setup.

\clearpage

\begin{table}[p]\centering
    \vskip-1cm
    \begin{tabular}{|c|c||c|c|c|c||c|c|c|c||c|}\hline
    Measurement& $\gamma$& $\alpha$& $\beta$& $\eta_a$& $\eta_b$& $r_{a_1}$& $r_{a_2}$& $r_{b_1}$& $r_{b_2}$& S \\ \hline\hline
on/off&$ 0.033$&$ 1.000$&$ 1.000$&$ 0.930$&$ 0.930$&$ 0.026$&$ 0.123$&$ 0.0001$&$ 0.225$&$ 2.023$\\\hline
on/off&$ 0.130$&$ 1.000$&$ 1.000$&$ 1.000$&$ 1.000$&$ 0.009$&$ 0.164$&$ 0.009$&$ 0.165$&$ 2.351$\\\hline
on/off&$ 0.715$&$ 1.000$&$ 1.000$&$ 1.000$&$ 1.000$&$ 0.008$&$ 0.102$&$ 0.005$&$ 0.176$&$ 2.027$\\\hline
on/off&$ 0.130$&$ 2.000$&$ 2.000$&$ 1.000$&$ 1.000$&$ 0.005$&$ 0.065$&$ 0.005$&$ 0.068$&$ 2.566$\\\hline
on/off&$ 0.163$&$ 2.000$&$ 2.000$&$ 0.870$&$ 0.870$&$ 0.007$&$ 0.078$&$ 0.007$&$ 0.082$&$ 2.001$\\\hline
on/off&$ 0.812$&$ 2.000$&$ 2.000$&$ 1.000$&$ 1.000$&$ 0.004$&$ 0.039$&$ 0.003$&$ 0.069$&$ 2.017$\\\hline
on/off&$ 0.033$&$ 3.000$&$ 3.000$&$ 0.860$&$ 0.860$&$ 0.012$&$ 0.030$&$ 0.0001$&$ 0.052$&$ 2.006$\\\hline
on/off&$ 0.163$&$ 3.000$&$ 3.000$&$ 1.000$&$ 1.000$&$ 0.003$&$ 0.032$&$ 0.002$&$ 0.033$&$ 2.624$\\\hline
on/off&$ 0.617$&$ 3.000$&$ 3.000$&$ 0.910$&$ 0.910$&$ 0.003$&$ 0.027$&$ 0.002$&$ 0.039$&$ 2.012$\\\hline
on/off&$ 0.065$&$ 6.000$&$ 6.000$&$ 0.850$&$ 0.850$&$ 0.003$&$ 0.008$&$ 0.0001$&$ 0.014$&$ 2.012$\\\hline
on/off&$ 0.098$&$ 6.000$&$ 6.000$&$ 0.850$&$ 0.850$&$ 0.003$&$ 0.008$&$ 0.0001$&$ 0.014$&$ 2.014$\\\hline
on/off&$ 0.098$&$ 6.000$&$ 6.000$&$ 1.000$&$ 1.000$&$ 0.0007$&$ 0.009$&$ 0.0008$&$ 0.008$&$ 2.671$\\\hline
on/off&$ 0.098$&$ 8.000$&$ 8.000$&$ 1.000$&$ 1.000$&$ 0.0004$&$ 0.005$&$ 0.0004$&$ 0.005$&$ 2.677$\\\hline
on/off&$ 0.130$&$ 8.000$&$ 8.000$&$ 0.840$&$ 0.840$&$ 0.0007$&$ 0.006$&$ 0.0007$&$ 0.007$&$ 2.003$\\\hline
on/off&$ 0.585$&$ 8.000$&$ 8.000$&$ 0.890$&$ 0.890$&$ 0.0005$&$ 0.004$&$ 0.0004$&$ 0.006$&$ 2.009$\\\hline
on/off&$ 0.098$&$ 10.000$&$ 10.000$&$ 1.000$&$ 1.000$&$ 0.0003$&$ 0.003$&$ 0.0003$&$ 0.003$&$ 2.680$\\\hline
on/off&$ 0.130$&$ 10.000$&$ 10.000$&$ 0.840$&$ 0.840$&$ 0.0004$&$ 0.004$&$ 0.0004$&$ 0.004$&$ 2.006$\\\hline
on/off&$ 0.552$&$ 10.000$&$ 10.000$&$ 0.880$&$ 0.880$&$ 0.0004$&$ 0.003$&$ 0.0003$&$ 0.004$&$ 2.001$\\\hline
parity&$ 0.029$&$ 1.000$&$ 1.000$&$ 0.960$&$ 0.960$&$ 0.002$&$ 0.088$&$ 0.002$&$ 0.089$&$ 2.007$\\\hline
parity&$ 2.000$&$ 1.000$&$ 1.000$&$ 0.990$&$ 0.990$&$ 0.0005$&$ 0.010$&$ 0.003$&$ 0.092$&$ 2.056$\\\hline
parity&$ 2.000$&$ 1.000$&$ 1.000$&$ 1.000$&$ 1.000$&$ 0.0005$&$ 0.010$&$ 0.003$&$ 0.091$&$ 2.249$\\\hline
parity&$ 0.029$&$ 2.000$&$ 2.000$&$ 0.950$&$ 0.950$&$ 0.0009$&$ 0.028$&$ 0.0009$&$ 0.028$&$ 2.011$\\\hline
parity&$ 2.000$&$ 2.000$&$ 2.000$&$ 0.990$&$ 0.990$&$ 0.0001$&$ 0.004$&$ 0.003$&$ 0.023$&$ 2.146$\\\hline
parity&$ 2.000$&$ 2.000$&$ 2.000$&$ 1.000$&$ 1.000$&$ 0.0001$&$ 0.004$&$ 0.003$&$ 0.023$&$ 2.346$\\\hline
parity&$ 0.029$&$ 3.000$&$ 3.000$&$ 0.950$&$ 0.950$&$ 0.0004$&$ 0.013$&$ 0.0004$&$ 0.013$&$ 2.021$\\\hline
parity&$ 2.000$&$ 3.000$&$ 3.000$&$ 0.990$&$ 0.990$&$ 0.0001$&$ 0.002$&$ 0.002$&$ 0.010$&$ 2.169$\\\hline
parity&$ 2.000$&$ 3.000$&$ 3.000$&$ 1.000$&$ 1.000$&$ 0.0001$&$ 0.002$&$ 0.002$&$ 0.010$&$ 2.371$\\\hline
parity&$ 0.457$&$ 6.000$&$ 6.000$&$ 0.940$&$ 0.940$&$ 0.0001$&$ 0.003$&$ 0.0001$&$ 0.004$&$ 2.000$\\\hline
parity&$ 2.000$&$ 6.000$&$ 6.000$&$ 0.990$&$ 0.990$&$ 0.0001$&$ 0.0006$&$ 0.0005$&$ 0.002$&$ 2.185$\\\hline
parity&$ 2.000$&$ 6.000$&$ 6.000$&$ 1.000$&$ 1.000$&$ 0.0001$&$ 0.0006$&$ 0.0005$&$ 0.002$&$ 2.388$\\\hline
parity&$ 0.429$&$ 8.000$&$ 8.000$&$ 0.940$&$ 0.940$&$ 0.0001$&$ 0.002$&$ 0.0001$&$ 0.002$&$ 2.000$\\\hline
parity&$ 2.000$&$ 8.000$&$ 8.000$&$ 0.980$&$ 0.980$&$ 0.0001$&$ 0.0004$&$ 0.0003$&$ 0.001$&$ 2.002$\\\hline
parity&$ 2.000$&$ 8.000$&$ 8.000$&$ 1.000$&$ 1.000$&$ 0.0001$&$ 0.0004$&$ 0.0003$&$ 0.001$&$ 2.391$\\\hline
parity&$ 0.400$&$ 10.000$&$ 10.000$&$ 0.940$&$ 0.940$&$ 0.0001$&$ 0.001$&$ 0.0001$&$ 0.001$&$ 2.000$\\\hline
parity&$ 2.000$&$ 10.000$&$ 10.000$&$ 0.980$&$ 0.980$&$ 0.0001$&$ 0.0002$&$ 0.0002$&$ 0.0008$&$ 2.003$\\\hline
parity&$ 2.000$&$ 10.000$&$ 10.000$&$ 1.000$&$ 1.000$&$ 0.0001$&$ 0.0002$&$ 0.0002$&$ 0.0008$&$ 2.392$\\\hline
    \end{tabular}
    \caption{Exemplary values of the CHSH parameter $S$ numerically optimized for on/off and parity measurements for different values of the coherent state amplitude $\gamma$, local oscillator amplitudes $\alpha,\beta$, efficiencies $\eta_{a,b}$, along with optimal Bell settings $r_{a_{1,2}},r_{b_{1,2}}$.}
    \label{tab:normal_regime}
\end{table}

\begin{table}[p]\centering
    \vskip-0.2cm
    \begin{tabular}{|c|c||c|c||c|c|c|c||c|}\hline
    Measurement& $\gamma$& $\eta_a$& $\eta_b$& $\delta_{\alpha_1}$& $\delta_{\alpha_2}$& $\delta_{\beta_1}$& $\delta_{\beta_2}$& S \\ \hline\hline
on/off&$ 0.325$&$ 0.827$&$ 0.827$&$-0.216$&$ 0.601$&$-0.197$&$ 0.633$&$ 2.001$\\\hline
on/off&$ 0.390$&$ 1.000$&$ 1.000$&$-0.183$&$ 0.559$&$-0.168$&$ 0.609$&$ 2.714$\\\hline
on/off&$ 1.235$&$ 0.991$&$ 0.991$&$-0.503$&$ 0.541$&$-0.030$&$ 1.231$&$ 2.001$\\\hline
parity&$ 0.300$&$ 0.941$&$ 0.941$&$-65.174$&$ 358.105$&$-65.808$&$ 333.049$&$ 2.001$\\\hline
parity&$ 2.000$&$ 0.982$&$ 0.982$&$-139.372$&$ 288.458$&$-16.114$&$ 154.916$&$ 2.037$\\\hline
parity&$ 2.000$&$ 1.000$&$ 1.000$&$-137.443$&$ 284.066$&$-15.599$&$ 152.203$&$ 2.394$\\\hline
    \end{tabular}
    \caption{Exemplary values of the CHSH parameter $S$ numerically optimized for on/off and parity measurements in the displacement regime, for different values of the coherent state amplitude $\gamma$, efficiencies $\eta_{a,b}$, along with optimal Bell settings $\delta_{\alpha_{1,2}},\delta_{\beta_{1,2}}$.}
    \label{tab:displacement_regime}
\end{table}

\clearpage

\begin{table}[p]\centering
    \vskip-1cm
    \begin{tabular}{|c|c||c|c|c|c||c|c|c|c|c|c||c|}\hline
    Measurement& $\gamma$& $\alpha$& $\beta$& $\eta_a$& $\eta_b$& $\theta_1$& $\phi_1$& $\theta_2$& $\phi_2$& $r_{b_1}$& $r_{b_2}$& S \\ \hline\hline
on/off&$ 0.033$&$ 1.000$&$ 1.000$&$ 0.880$&$ 0.880$&$ 0.314$&$-1.195$&$ 1.650$&$ 1.604$&$ 0.018$&$ 0.196$&$ 2.009$\\\hline
on/off&$ 0.098$&$ 1.000$&$ 1.000$&$ 1.000$&$ 1.000$&$ 0.068$&$-1.502$&$ 1.865$&$ 1.575$&$ 0.064$&$ 0.089$&$ 2.562$\\\hline
on/off&$ 0.812$&$ 1.000$&$ 1.000$&$ 1.000$&$ 1.000$&$ 0.012$&$-1.584$&$ 4.882$&$ 1.568$&$ 0.048$&$ 0.045$&$ 2.022$\\\hline
on/off&$ 0.098$&$ 2.000$&$ 2.000$&$ 0.840$&$ 0.840$&$ 0.017$&$-1.586$&$ 17.352$&$ 1.569$&$ 0.023$&$ 0.022$&$ 2.034$\\\hline
on/off&$ 0.098$&$ 2.000$&$ 2.000$&$ 1.000$&$ 1.000$&$ 0.028$&$-1.563$&$ 20.769$&$ 1.566$&$ 0.014$&$ 0.015$&$ 2.721$\\\hline
on/off&$ 0.845$&$ 2.000$&$ 2.000$&$ 0.990$&$ 0.990$&$ 0.003$&$-1.572$&$  17340$&$ 1.571$&$ 0.009$&$ 0.009$&$ 2.020$\\\hline
on/off&$ 0.065$&$ 3.000$&$ 3.000$&$ 0.820$&$ 0.820$&$-0.005$&$-1.575$&$ 2.934$&$ 1.579$&$ 0.024$&$ 0.025$&$ 2.001$\\\hline
on/off&$ 0.065$&$ 3.000$&$ 3.000$&$ 1.000$&$ 1.000$&$-0.007$&$-1.574$&$ 4.329$&$ 1.570$&$ 0.015$&$ 0.015$&$ 2.762$\\\hline
on/off&$ 0.878$&$ 3.000$&$ 3.000$&$ 1.000$&$ 1.000$&$-0.003$&$-1.571$&$  38436$&$ 1.571$&$ 0.008$&$ 0.008$&$ 2.005$\\\hline
on/off&$ 0.033$&$ 6.000$&$ 6.000$&$ 1.000$&$ 1.000$&$-0.008$&$-1.578$&$ 1.146$&$ 1.593$&$ 0.004$&$ 0.004$&$ 2.790$\\\hline
on/off&$ 0.065$&$ 6.000$&$ 6.000$&$ 0.820$&$ 0.820$&$ 0.004$&$-1.577$&$ 4.180$&$ 1.572$&$ 0.007$&$ 0.007$&$ 2.033$\\\hline
on/off&$ 0.878$&$ 6.000$&$ 6.000$&$ 1.000$&$ 1.000$&$ 0.001$&$-1.573$&$   4126$&$ 1.571$&$ 0.002$&$ 0.002$&$ 2.020$\\\hline
on/off&$ 0.033$&$ 8.000$&$ 8.000$&$ 1.000$&$ 1.000$&$ 0.004$&$ 1.555$&$ 1.624$&$-1.562$&$ 0.002$&$ 0.002$&$ 2.793$\\\hline
on/off&$ 0.065$&$ 8.000$&$ 8.000$&$ 0.820$&$ 0.820$&$ 0.012$&$ 1.577$&$-2.730$&$-1.571$&$ 0.004$&$ 0.004$&$ 2.038$\\\hline
on/off&$ 0.878$&$ 8.000$&$ 8.000$&$ 1.000$&$ 1.000$&$-0.003$&$ 1.570$&$   1632$&$-1.569$&$ 0.001$&$ 0.001$&$ 2.022$\\\hline
on/off&$ 0.033$&$ 10.000$&$ 10.000$&$ 0.810$&$ 0.810$&$-0.052$&$ 1.610$&$ 1.429$&$-1.562$&$ 0.003$&$ 0.002$&$ 2.001$\\\hline
on/off&$ 0.033$&$ 10.000$&$ 10.000$&$ 1.000$&$ 1.000$&$-0.068$&$ 1.627$&$ 1.952$&$-1.575$&$ 0.002$&$ 0.001$&$ 2.792$\\\hline
on/off&$ 0.878$&$ 10.000$&$ 10.000$&$ 1.000$&$ 1.000$&$-0.0001$&$ 1.570$&$-56.450$&$-1.572$&$ 0.0008$&$ 0.0008$&$ 2.023$\\\hline
parity&$ 0.371$&$ 1.000$&$ 1.000$&$ 0.910$&$ 0.910$&$ 0.0001$&$ 1.571$&$-4.746$&$-1.571$&$ 0.034$&$ 0.034$&$ 2.000$\\\hline
parity&$ 2.000$&$ 1.000$&$ 1.000$&$ 0.970$&$ 0.970$&$ 0.0001$&$ 1.571$&$-0.0001$&$-1.571$&$ 0.006$&$ 0.006$&$ 2.013$\\\hline
parity&$ 2.000$&$ 1.000$&$ 1.000$&$ 1.000$&$ 1.000$&$ 0.0001$&$ 1.571$&$-0.0001$&$-1.571$&$ 0.006$&$ 0.006$&$ 2.627$\\\hline
parity&$ 0.229$&$ 2.000$&$ 2.000$&$ 0.900$&$ 0.900$&$ 0.0001$&$ 1.571$&$-1.351$&$-1.571$&$ 0.006$&$ 0.006$&$ 2.001$\\\hline
parity&$ 2.000$&$ 2.000$&$ 2.000$&$ 0.970$&$ 0.970$&$ 0.0001$&$ 1.571$&$-0.015$&$-1.571$&$ 0.0010$&$ 0.0010$&$ 2.098$\\\hline
parity&$ 2.000$&$ 2.000$&$ 2.000$&$ 1.000$&$ 1.000$&$ 0.0001$&$ 1.571$&$-0.017$&$-1.571$&$ 0.0009$&$ 0.0009$&$ 2.733$\\\hline
parity&$ 0.200$&$ 3.000$&$ 3.000$&$ 0.900$&$ 0.900$&$ 0.0001$&$ 1.571$&$-0.376$&$-1.571$&$ 0.006$&$ 0.006$&$ 2.011$\\\hline
parity&$ 2.000$&$ 3.000$&$ 3.000$&$ 0.970$&$ 0.970$&$ 0.0001$&$ 1.571$&$-0.015$&$-1.571$&$ 0.0010$&$ 0.0010$&$ 2.117$\\\hline
parity&$ 2.000$&$ 3.000$&$ 3.000$&$ 1.000$&$ 1.000$&$ 0.0001$&$ 1.571$&$-0.017$&$-1.571$&$ 0.0009$&$ 0.0009$&$ 2.757$\\\hline
parity&$ 0.457$&$ 6.000$&$ 6.000$&$ 0.890$&$ 0.890$&$-0.0001$&$ 1.571$&$ 1.497$&$-1.571$&$ 0.001$&$ 0.001$&$ 2.001$\\\hline
parity&$ 2.000$&$ 6.000$&$ 6.000$&$ 0.970$&$ 0.970$&$-0.0001$&$ 1.571$&$-11.894$&$-1.571$&$ 0.0003$&$ 0.0003$&$ 2.130$\\\hline
parity&$ 2.000$&$ 6.000$&$ 6.000$&$ 1.000$&$ 1.000$&$-0.0001$&$ 1.571$&$-12.190$&$-1.571$&$ 0.0002$&$ 0.0002$&$ 2.772$\\\hline
parity&$ 0.429$&$ 8.000$&$ 8.000$&$ 0.890$&$ 0.890$&$ 0.0001$&$ 1.571$&$-0.556$&$-1.571$&$ 0.0008$&$ 0.0008$&$ 2.000$\\\hline
parity&$ 2.000$&$ 8.000$&$ 8.000$&$ 0.970$&$ 0.970$&$ 0.0001$&$ 1.571$&$-15.943$&$-1.571$&$ 0.0001$&$ 0.0001$&$ 2.131$\\\hline
parity&$ 2.000$&$ 8.000$&$ 8.000$&$ 1.000$&$ 1.000$&$ 0.0001$&$ 1.571$&$-17.521$&$-1.571$&$ 0.0001$&$ 0.0001$&$ 2.775$\\\hline
parity&$ 0.429$&$ 10.000$&$ 10.000$&$ 0.890$&$ 0.890$&$ 0.0001$&$ 1.571$&$ 0.376$&$-1.571$&$ 0.0005$&$ 0.0005$&$ 2.001$\\\hline
parity&$ 2.000$&$ 10.000$&$ 10.000$&$ 0.970$&$ 0.970$&$ 0.0001$&$ 1.571$&$ 0.686$&$-1.571$&$ 0.0001$&$ 0.0001$&$ 2.132$\\\hline
parity&$ 2.000$&$ 10.000$&$ 10.000$&$ 1.000$&$ 1.000$&$ 0.0001$&$ 1.571$&$ 0.759$&$-1.571$&$ 0.0001$&$ 0.0001$&$ 2.776$\\\hline
    \end{tabular}
    \caption{Exemplary values of the CHSH parameter $S$ numerically optimized for on/off and parity measurements in hybrid mode for different values of the coherent state amplitude $\gamma$, local oscillator amplitudes $\alpha,\beta$, efficiencies $\eta_{a,b}$, along with optimal Bell settings $\theta_{1,2},\phi_{1,2},r_{b_{1,2}}$.}
    \label{tab:hybrid_normal_regime}
\end{table}

\begin{table}[p]\centering
    \vskip-0.2cm
    \begin{tabular}{|c|c||c|c||c|c|c|c|c|c||c|}\hline
    Measurement& $\gamma$& $\eta_a$& $\eta_b$& $\theta_1$& $\phi_1$& $\theta_2$& $\phi_2$& $\delta_{\beta_1}$& $\delta_{\beta_2}$& S \\ \hline\hline
on/off&$ 0.390$&$ 0.795$&$ 0.795$&$-0.0001$&$ 1.571$&$ 3.275$&$ 0.0001$&$-0.487$&$ 0.487$&$ 2.004$\\\hline
on/off&$ 0.423$&$ 1.000$&$ 1.000$&$-0.0001$&$ 1.571$&$ 3.280$&$ 0.0001$&$-0.417$&$ 0.417$&$ 2.828$\\\hline
on/off&$ 1.300$&$ 0.905$&$ 0.905$&$-1.189$&$ 1.853$&$ 3.142$&$-0.0006$&$-1.238$&$ 0.065$&$ 2.009$\\\hline
parity&$ 0.400$&$ 0.891$&$ 0.891$&$-0.0001$&$ 1.571$&$ 3.146$&$-0.0001$&$-0.224$&$ 0.224$&$ 2.003$\\\hline
parity&$ 2.000$&$ 0.964$&$ 0.964$&$ 0.511$&$ 2.637$&$ 56.549$&$ 0.001$&$-1.618$&$-0.0003$&$ 2.018$\\\hline
parity&$ 2.000$&$ 1.000$&$ 1.000$&$ 0.463$&$ 2.677$&$ 56.549$&$ 0.001$&$-1.633$&$-0.0003$&$ 2.778$\\\hline
    \end{tabular}
    \caption{Exemplary values of the CHSH parameter $S$ numerically optimized for on/off and parity measurements in the displacement regime in hybrid mode, for different values of the coherent state amplitude $\gamma$, efficiencies $\eta_{a,b}$, along with optimal Bell settings $\theta_{1,2},\phi_{1,2},\delta_{\beta_{1,2}}$.}
    \label{tab:hybrid_displacement_regime}
\end{table}

\clearpage

\section{HYBRID MEASUREMENTS AND CHSH RIGIDITY THEOREM}

Consider the hybrid version of the zero/non-zero test, in which Alice performs arbitrary spin/polarization measurements defined by the Hermitian operators \((\sin \theta \cos \phi)X + (\sin \theta \sin \phi)Y + (\cos \theta)Z\)

\begin{align}
A_1 = \begin{pmatrix}
\cos \theta_1  & e^{-i\phi_1}\sin \theta_1  \\
e^{i\phi_1} \sin \theta_1  & -\cos \theta_1
\end{pmatrix} 
\hspace{3mm},\hspace{3mm}
A_2 = \begin{pmatrix}
\cos \theta_2  & e^{-i\phi_2}\sin \theta_2  \\
e^{i\phi_2}\sin \theta_2  & -\cos \theta_2
\end{pmatrix},
\end{align}
while Bob performs the zero/non-zero measurement as before, with operators
\begin{align}
B_1 =  2|\delta_\beta_1\rangle\langle\delta_\beta_1| - \mathbb{1}
\hspace{3mm},\hspace{3mm}
B_2 =  2|\delta_\beta_2\rangle\langle\delta_\beta_2| - \mathbb{1}.
\end{align}

The state $\ket{\Psi} = \frac{1}{\sqrt{2}}\bigl(\ket{0}\ket{\text{cat}^-} + \ket{1}\ket{\text{cat}^+}\bigr)$ has the same form as a maximally entangled Bell pair, with the states $|\text{cat}^-\rangle$ and $|\text{cat}^+\rangle$ playing the role of macroscopic qubit states. Therefore, if Alice and Bob could both perform arbitrary qubit measurements, they could maximally violate the CHSH inequality, using the canonical measurements $A_1 = -Z, \hspace{1mm} A_2 = -X, \hspace{1mm} B_1 = \frac{-1}{\sqrt{2}}(Z + X)$, and  $B_2 = \frac{-1}{\sqrt{2}}(Z - X)$. However, we are considering a hybrid approach where Bob is restricted only to feasible measurements on his macroscopic photonic system, namely displacement operations and photon number measurements. Nevertheless, we show that maximal violation can still be achieved, since for one specific value of $\gamma$, Bob's measurements $B_1$ and $B_2$ are equivalent to the canonical qubit measurements in the basis of $|\text{cat}^\pm\rangle$.

From our numerical optimization, we found that Bob's optimal displacements are $\delta\beta_1 = -\gamma$ and $\delta\beta_2 = +\gamma$, exactly canceling the coherent state amplitude of the cat state. Taking this as our starting point, by writing

\begin{equation}
|\pm\gamma\rangle = \frac{N_+|\text{cat}+\rangle \pm N_-|\text{cat}-\rangle}{2}
\end{equation}

We can express $B_1$ and $B_2$ in the basis of $|\text{cat} \pm\rangle$:

\begin{align}
B_1 &= 2|\text{-}\gamma\rangle\langle\text{-}\gamma| - \mathbb{1} = \frac{1}{2}(N_+|\text{cat}^+\rangle - N_-|\text{cat}^-\rangle)(N_+\langle\text{cat}^+| - N_-\langle\text{cat}^-|) - \mathbb{1} \nonumber\\
&= \frac{1}{2}(N_-^2|\text{cat}^-\rangle\langle\text{cat}^-| - N_+N_-|\text{cat}^-\rangle\langle\text{cat}^+| - N_+N_-|\text{cat}^+\rangle\langle\text{cat}^-| + N_+^2|\text{cat}^+\rangle\langle\text{cat}^+|) - \mathbb{1} .
\end{align}

Here, $ \mathbb{1}$ is the identity operator, and to express it in terms of $|\text{cat}^\pm\rangle$, we need to introduce a full set of orthonormal states $|e_i\rangle$. We have two already $|e_0\rangle = |\text{cat}^-\rangle$ and $|e_1\rangle = |\text{cat}^+\rangle$, but these do not completely span the space (we are considering only one specific value of $\gamma$). For example, the single photon state $|1\rangle$ can't be expressed as a linear combination of these two states. To complete the basis, we need an arbitrary set of additional states $|e_i\rangle$ with $i \geq 2$ which are each mutually orthonormal and also orthonormal to $|\text{cat}^ \pm\rangle$. The identity operator can then be written as

\begin{equation}
 \mathbb{1} = |\text{cat}^- \rangle \langle \text{cat}^-| + |\text{cat}^+ \rangle \langle \text{cat}^+| + \sum^\infty_{i=2} |e_i \rangle\langle e_i| ,
\end{equation}

and we obtain

\begin{equation}
B_1 = \left(\frac{N^2_-}{2}-1\right) |cat^-\rangle \langle cat^-| - \frac{N_+N_-}{2} |cat^-\rangle \langle cat^+| - \frac{N_+N_-}{2} |cat^+\rangle \langle cat^-| + \left(\frac{N^2_+}{2}-1\right) |cat^+\rangle \langle cat^+| - \sum^\infty_{i=2}|e_i\rangle \langle e_i|
\end{equation}
In our orthonormal basis, $B_1$ can be represented as the matrix

\begin{equation}
B_1 =
\begin{pmatrix}
\frac{N^2_-}{2}-1 & -\frac{N_+N_-}{2}  \\
-\frac{N_+N_-}{2} & \frac{N^2_+}{2}-1  \\
 & & -1 \\
 & & & \ddots 
\end{pmatrix} =  
\begin{pmatrix}
-e^{-2\gamma^2} & -\sqrt{1 - e^{-4\gamma^2}}\\
-\sqrt{1 - e^{-4\gamma^2}} & e^{-2\gamma^2}  \\
 & & -1 \\
 & & & \ddots
\end{pmatrix}.
\end{equation}

Similarly, $B_2 = 2|\gamma\rangle \langle\gamma| - \mathbb{1}$ is

\begin{equation}
B_2 =
\begin{pmatrix}
\frac{N^2_-}{2}-1 & \frac{N_+N_-}{2}  \\
\frac{N_+N_-}{2} & \frac{N^2_+}{2}-1  \\
 & & -1 \\
 & & & \ddots 
\end{pmatrix} =  
\begin{pmatrix}
-e^{-2\gamma^2} & \sqrt{1 - e^{-4\gamma^2}}\\
\sqrt{1 - e^{-4\gamma^2}} & e^{-2\gamma^2}  \\
 & & -1 \\
 & & & \ddots
\end{pmatrix}.
\end{equation}

Then, at the specific point $\gamma = \pm\sqrt{\ln(2)}/{2} = \pm 0.416$, these act as the canonical measurements $B_1 = \frac{-1}{\sqrt{2}}(Z + X)$, $B_2 = \frac{-1}{\sqrt{2}}(Z - X)$ in the upper-left $2 \times 2$ subspace of $|\text{cat}^\pm\rangle$. Thus, we obtain the maximal violation of the CHSH inequality $S = 2\sqrt{2}$.

We can perform a similar analysis for the hybrid even/odd test, here Alice again uses general qubit measurements, but Bob measures the Hermitian operators

\begin{align}
B1 &= \sum_{n=0}^{\infty} (-1)^{n} D(\delta_{\beta_1}) |n\rangle \langle n| D(-\delta_{\beta_1}), \\
B2 &= \sum_{n=0}^{\infty} (-1)^{n} D(\delta_{\beta_2}) |n\rangle \langle n| D(-\delta_{\beta_2}).
\end{align}

From the numerical optimization, we know that the maximal violation occurs in the limit of large gamma $\gamma \to \infty$ while $\delta$ remains finite. Considering this limit, where $|\text{cat}^\pm\rangle = \frac{1}{\sqrt{2}}(|\gamma\rangle \pm |-\gamma\rangle)$, we wish to show that $B_1$ and $B_2$ act as the canonical measurements in the $|\text{cat} ^\pm\rangle$ basis, up to a general rotation. For a general displacement $\delta$, we can see how Bob's measurement $B_i$ acts in this subspace

\begin{align}
\langle\text{cat}^-| B_i|\text{cat}^-\rangle &= \frac{1}{2}(\langle\gamma| B_i|\gamma\rangle - \langle\gamma| B_i|-\gamma\rangle - \langle -\gamma| B_i|\gamma\rangle + \langle -\gamma| B_i|-\gamma\rangle), \\
\langle\text{cat}^-| B_i|\text{cat}^+\rangle &= \frac{1}{2}(\langle\gamma| B_i|\gamma\rangle + \langle\gamma| B_i|-\gamma\rangle - \langle -\gamma| B_i|\gamma\rangle - \langle -\gamma| B_i|-\gamma\rangle), \\
\langle\text{cat}^+| B_i|\text{cat}^-\rangle &= \frac{1}{2}(\langle\gamma| B_i|\gamma\rangle - \langle\gamma| B_i|-\gamma\rangle + \langle - \gamma| B_i|\gamma\rangle - \langle -\gamma| B_i| - \gamma\rangle), \\
\langle\text{cat}^+| B_i|\text{cat}^+\rangle &= \frac{1}{2}(\langle\gamma| B_i|\gamma\rangle + \langle\gamma| B_i| - \gamma\rangle + \langle - \gamma| B_i|\gamma\rangle + \langle - \gamma| B_i|-\gamma\rangle).
\end{align}

Replacing lossless condition i.e. $\eta_B=1$ in \ref{OpticalParityCPart} gives us the terms $\langle\gamma| B_i|\gamma\rangle = \exp\left[-2|\gamma - \delta|^2\right]$ and $\langle\gamma| B_i|-\gamma\rangle = \exp\left[-2|\delta|^2 + 4i\, \text{Im}(\gamma^*\delta)\right]$ where $\text{Im}(x)$ denotes the imaginary component of $x$. From our assumptions that $\gamma \to \infty$ and $\delta$ remains finite, the former simplifies to $\langle\gamma| B_i|\gamma\rangle = 0$. Thus we obtain

\begin{align}
\langle\text{cat}^+| B_i|\text{cat}^+\rangle &= +e^{-2|\delta|^2}\cos\left[4\,\text{Im}(\gamma^*\delta)\right],\nonumber\\
\langle\text{cat}^+| B_i|\text{cat}^-\rangle &= -ie^{-2|\delta|^2}\sin\left[4\,\text{Im}(\gamma^*\delta)\right],\nonumber\\
\langle\text{cat}^-| B_i|\text{cat}^+\rangle &= +ie^{-2|\delta|^2}\sin\left[4\,\text{Im}(\gamma^*\delta)\right],\nonumber\\
\langle\text{cat}^-| B_i|\text{cat}^-\rangle &= -e^{-2|\delta|^2}\cos\left[4\,\text{Im}(\gamma^*\delta)\right].\nonumber
\end{align}

or in matrix notation \\

\begin{equation}
B_i =
\begin{pmatrix}
e^{-2|\delta|^2}\cos[4\,\text{Im}(\gamma^*\delta)] & -ie^{-2|\delta|^2}\sin[4\,\text{Im}(\gamma^*\delta)] \\
ie^{-2|\delta|^2}\sin[4\,\text{Im}(\gamma^*\delta)] & -e^{-2|\delta|^2}\cos[4\,\text{Im}(\gamma^*\delta)]
\end{pmatrix}
\end{equation}

Then we can observe that by choosing $\delta = \pm{i\pi}/{16\gamma}$, we can obtain $B_1 = \frac{1}{\sqrt{2}}(Z + Y)$ and $B_2 = \frac{1}{\sqrt{2}}(Z - Y)$, which are the canonical measurements up to a rotation in the X-Y plane. If Alice performs an identical rotation, measuring $A_1 = Z$ and $A_2 = Y$, we once again maximally violate the CHSH inequality $S = 2\sqrt{2}$.